\documentclass[fleqn,usenatbib]{mnras}

\usepackage{newtxtext,newtxmath}
\usepackage[T1]{fontenc}
\usepackage{ae,aecompl}
\usepackage{xcolor}

\usepackage{graphicx}	% Including figure files
\usepackage{amsmath}	% Advanced maths commands

\def\surfs{{\sc surfs}}
\def\shark{{\sc Shark}}
\title[Star-forming galaxies at $z>10$ in \shark]{Bright star-forming galaxies naturally forming at $z>10$ in the \shark\ semi-analytic model}

\author[C.~D.~P. Lagos et al.]{
\parbox[t]{\textwidth}{
\vspace{-0.5cm}
Claudia del P. Lagos$^{1,2}$\thanks{E-mail: claudia.lagos@icrar.org}, {\'A}ngel Chandro-G{\'o}mez${^1}$, Chris Power$^{1}$, Aaron S.G. Robotham$^1$}
\vspace*{6pt} \\
$^{1}$International Centre for Radio Astronomy Research (ICRAR), M468, University of Western Australia, 35 Stirling Hwy, Crawley, \\WA 6009, Australia.\\
$^{2}$Cosmic Dawn Center (DAWN), Denmark.\\
\vspace*{-0.5cm}}

\date{Accepted XXX. Received YYY; in original form ZZZ}

\pubyear{2024}

\begin{document}
\label{firstpage}
\pagerange{\pageref{firstpage}--\pageref{lastpage}}
\maketitle

\begin{abstract}
The {\it James Webb Space Telescope} ({\it JWST}) has unveiled the existence of numerous $z>10$ bright ultraviolet (UV) galaxies, potentially challenging galaxy formation models in a $\Lambda$ Cold Dark Matter ($\Lambda$CDM) universe. Modifications to star formation and stellar feedback models have been suggested to alleviate the tension. However, the fundamental challenge is to design a galaxy formation model that simultaneously reproduces observations from $z=0$ to those at the highest redshifts. Here, we present predictions from the \shark\ semi-analytic model of galaxy formation, which is tuned to reproduce the $z=0$ universe. We show that the same model is capable of reproducing the current UV luminosity function constraints even up to $z=17$ without the need to invoke variations in the baryon physics model. This model is also capable of reproducing reasonably well the stellar mass function evolution from $z=0$ to $z=10$ and the cosmic star formation rate (SFR) density at $0\le z\le 15$, again with the same baryon physics and parameters. The key to the success of \shark\ is the triggering of starbursts by violent disk instabilities included in the model, which leads to an increase in the variance of the UV luminosity with increasing halo mass at $z>10$. We demonstrate that without this mechanism, galaxies are not bursty enough in the model to reproduce the observations at $z>10$. 
\end{abstract}


%%%%%%%%%%%%%%%%%%%%%%%%%%%%%%%%%%%%%%%%%%%%%%%%%%

%%%%%%%%%%%%%%%%% BODY OF PAPER %%%%%%%%%%%%%%%%%%

\section{Introduction}

%P1: how the JWST is allowing the discovery of galaxies at cosmic dawn, and the measurement of statistical properties of the galaxy population such as the UV LF up to z=20.

%P2: The challenges this has presented, with many models struggling to reproduce these observations. Recognise here though that GALFORM seems to behave pretty well. 

%P3: what solution have been offered: feedback-free starbursts, increased SF efficiency, and increased burstiness. Are the issues baryon physics, or is there anything more fundamental with the cosmological model?

%P4: what the real challenge is, is to be able to explain observations at vastly different cosmic times within the same galaxy formation theory framework. Somerville+25 showed that in extreme changes to the star formation model it's possible to reproduce the observations, but they do that at the expense of matching the lower redshift universe.

%P5: where Shark comes in. What the model has been able to reproduce (up to z=10) and why it's natural to test whether the same physics is capable of reproduce higher z observations.

The launch of the {\it James Webb Space Telescope} ({\it JWST}) has transformed our view of the earliest phases of galaxy formation, extending observations to within only a few hundred million years of the Big Bang. Deep imaging surveys, complemented by an increasing number of spectroscopic confirmations, have revealed substantial populations of galaxies out to redshifts $z \gtrsim 14$ (e.g. \citealt{Atek23, Carniani24, Harikane24, Schouws25, Naidu26}). For the first time, these observations have enabled the statistical characterisation of galaxy populations during cosmic dawn, including measurements of the rest-frame ultraviolet (UV) luminosity function (UVLF) over $6\lesssim z \lesssim 15$ (e.g. \citealt{Bouwens23,Donnan23,Donnan24,Finkelstein24,Harikane23,Harikane24,Casey24,Robertson24}). Because rest-frame UV emission traces recent star formation, the UVLF provides one of the most direct observational probes of the assembly of the first galaxies and therefore constitutes a fundamental benchmark for theories of galaxy formation and cosmic reionisation \citep{Robertson15,Madau14}.
%(e.g. Robertson et al. 2015; Madau & Dickinson 2014).

One of the most surprising early JWST results was the apparently high abundance of UV luminous galaxies at $z \ge 10$. Comparisons with many theoretical models developed prior to the {\it JWST} suggested that these galaxies were significantly more numerous than expected, leading to suggestions that either galaxy formation in the early Universe was substantially more efficient than previously believed or, more speculatively, that the observations might challenge aspects of the standard $\Lambda$ Cold Dark Matter ($\Lambda$CDM) paradigm itself (e.g. \citealt{Boylan-Kolchin23,Labbe23,Lovell23}). As larger surveys have become available, photometric redshifts have improved, and spectroscopic confirmations have accumulated, the discussion has shifted away from questioning the cosmological model towards identifying which aspects of baryonic physics require revision (e.g. \citealt{Yung24,Finkelstein24}). Nevertheless, most cosmological hydrodynamical simulations and semi-analytic models calibrated to reproduce galaxy populations at low redshift continue to underpredict the abundance of the brightest UV galaxies beyond $z\sim 10$ (e.g. \citealt{Kannan23,Yung24,Lu26}). An important exception is the Durham semi-analytic model GALFORM, whose pre-{\it JWST} predictions were shown to reproduce the observed UVLF remarkably well up to $z\sim 12$, and, after accounting for delayed dust grain growth, out to $z\approx 14$ (\citealt{Cowley18,Lu25}). These results demonstrate that the current observations do not necessarily require departures from $\Lambda$CDM, but instead place stringent constraints on the baryonic processes governing the formation of the earliest galaxies. 

The discrepancy between many theoretical predictions and the observed UVLF has motivated a wide range of proposed physical explanations. Several studies have argued that star formation may proceed considerably more efficiently in the dense interstellar media of early galaxies than is inferred from observations of nearby systems. For example, density-dependent star formation prescriptions motivated by cloud-scale simulations naturally increase the efficiency of converting gas into stars in high-pressure environments and substantially enhance the abundance of UV-bright galaxies (e.g. \citealt{Mauerhofer25, Somerville25}). Others have suggested weaker stellar feedback, allowing a larger fraction of baryons to remain available for star formation \citep{Dekel23,Li24}, while increasingly bursty star formation histories have also been shown to boost UV luminosities relative to models with smoother star formation \citep{Mason23,Wang24,Gelli24,Katz25,Semenov25}. Alternative explanations include evolving dust attenuation (e.g. \citealt{Ferrara23,Yung24}), the contribution of active galactic nuclei (AGN; e.g. \citealt{Pacucci22,Hegde24}), or variations of the stellar initial mass function (IMF; e.g. \citealt{Yung24,Mauerhofer25,Fontanot26,Lu25,Katz25,Durrant26}), with top-heavy or otherwise non-universal IMFs increasing the UV luminosity produced per unit stellar mass formed. Collectively, these studies demonstrate that the observed UVLF can be reproduced through a variety of modifications to baryonic physics, without necessarily invoking changes to the underlying cosmological framework.

However, reproducing the UVLF at $z>10$ alone is not a sufficient test of a galaxy formation model. Modern semi-analytic models and cosmological hydrodynamical simulations are typically calibrated to reproduce the well-characterised galaxy population at low redshift, including stellar mass functions, gas scaling relations, star formation histories and galaxy structural properties (see reviews of \citealt{Somerville14,Crain23}). A successful theory of galaxy formation must therefore simultaneously explain observations spanning more than 13 billion years of cosmic evolution using a single, physically consistent description of baryonic processes. Modifications introduced solely to match the earliest galaxies risk degrading the agreement achieved at later epochs. This difficulty has recently been illustrated by \citet{Somerville25}, who showed that sufficiently large increases in the star formation efficiency can reproduce the observed abundance of luminous galaxies during cosmic dawn, but at the expense of significantly overproducing the galaxy population at lower redshifts. The key challenge is therefore not simply to explain the brightest {\it JWST} galaxies, but to do so without sacrificing the successful description of galaxy evolution across the rest of cosmic history.

In this context, the \shark\ semi-analytic model provides an ideal framework to investigate whether the observed UVLF can emerge naturally from a galaxy formation model calibrated independently of the {\it JWST} results. \shark\ is an open-source semi-analytic model that incorporates physically motivated treatments of gas accretion and cooling, star formation, stellar feedback, black hole growth, AGN feedback, environmental processes and galaxy structural evolution \citep{Lagos18c,Lagos24}. The latest version of \shark\ \citep{Lagos24} has been calibrated exclusively to reproduce the stellar mass function in the local Universe, while successfully reproducing a broad range of independent observations, including gas scaling relations, the evolution of the stellar mass function to $z\approx 10$, the cosmic star formation history, the emergence of massive quenched galaxies in the early Universe, and the evolution of supermassive black holes \citep{Lagos18c,Lagos24,Bravo25,Oxland26}. None of these observational constraints includes the high-redshift UVLF measured by the {\it JWST}, making its prediction a genuine test of the underlying physical model rather than a consequence of parameter tuning.

In this paper, we investigate the evolution of the UVLF predicted by \shark\ over the redshift range $6\le z\le 17$. We show that the fiducial model reproduces the currently observed UVLF across this entire redshift interval without requiring modifications to the baryonic physics or recalibration of the model parameters, and demonstrate that this success is driven by the increasing stochasticity of star formation associated with violent disk instabilities, which trigger compact, efficient starbursts at early cosmic times. We further place these results in the broader context of galaxy evolution by showing that the same model provides a good description of the evolution of the stellar mass function and cosmic SFR density from the present day to $z\sim15$. Our results therefore suggest that the abundance of UV-bright galaxies during cosmic dawn can arise naturally within a $\Lambda$CDM framework when sufficiently bursty modes of star formation are captured. The paper is organised as follows. \S~\ref{sams} describes \shark, the physical processes relevant for the UV emission of galaxies, and the suite of $N$-body simulations used in this work. \S~\ref{uvlfsec} presents the predicted UVLF at $6\le z\le17$ and explores the effects of dust attenuation, variations of the fiducial model, and the origin of the evolving star formation stochasticity. \S~\ref{allcosmicepoch} places these predictions in the context of the stellar mass function and cosmic SFR density across cosmic time.  Finally, \S~\ref{discussion} and \ref{conclusions} present a discussion of our results and our main conclusions, respectively.  

\section{The semi-analytic model \shark}\label{sams}

\shark\ is an open-source semi-analytic model of galaxy formation was first 
introduced in \citet{Lagos18}, and is publicly available on GitHub\footnote{\href{https://github.com/ICRAR/shark}{\url{https://github.com/ICRAR/shark}}}. The latest released version  is \shark\ v2.0, which was introduced in \citet{Lagos24}. 

The model includes all the
physical processes that we think govern the formation and evolution
of galaxies. These are: (i) the collapse and merging of DM haloes;
(ii) the accretion of gas on to haloes, which is modulated by the
DM accretion rate; (iii) the shock heating and radiative cooling of
gas inside DM haloes, leading to the formation of galactic disks
via conservation of specific angular momentum of the cooling gas;
(iv) star formation in galaxy disks; (v) stellar feedback from the
evolving stellar populations; (vi) chemical enrichment of stars and
gas; (vii) the growth of black holes (BHs) via gas accretion and merging with other BHs; (viii) feedback from Active Galactic Nuclei (AGN) in the form of outflows and jets; (ix) photoionization of the
intergalactic medium; (x) galaxy mergers driven by dynamical
friction within common DM haloes, which can trigger starbursts and
the formation and/or growth of spheroids; (xi) collapse of globally
unstable disks that also lead to starbursts and the formation and/or
growth of bulges; (xii) and environmental processes that affect the capability of satellite galaxies to retain or regrow their gas reservoir. \shark\ adopts a universal \citet{Chabrier03} initial mass function (IMF).

In this paper we use the latest version of \shark, which was introduced in \citet{Lagos24}. 
This new version represents a comprehensive improvement over the original version published in 2018, which 
included significant improvements in the modelling of AGN 
feedback to include two feedback modes (a core-jet powered one and radiation-pressure driven winds); 
environmental effects (tidal and ram pressure stripping); and the angular 
momentum exchange between baryon components. The AGN feedback model was shown to be critical 
to obtain a high abundance of massive-quenched galaxies at $z\ge 2$ \citep{Lagos24, Lagos25} that are 
comparable to the latest JWST measurements (e.g. \citealt{Baker25,Zhang26,Yang26}). \citet{Bravo25} also show
that this version of \shark\ predicts an evolving BH-stellar mass relation, so that BHs tend to be more 
massive at fixed stellar mass in the early Universe compared with $z=0$ galaxies, in broad agreement 
with {\it JWST} observations (see also \citealt{Izquierdo-Villalba26} for a large comparison of simulations with observations of BHs, including \shark). 

An important assumption in \shark\ and any SAM is that galaxies
can be described as having two structural components: a disk plus bulge, at any time. The main
distinction between these two components is their origin. Disks 
form stars from gas that is accreted on to the galaxy from
the halo or other galaxies through mergers, while bulges are built by stars that are accreted from satellite
galaxies and starbursts that are driven by galaxy mergers or disk
instabilities.

The free parameters of \shark\ were calibrated using an automatic optimiser that fits the $z\approx 0.1$ stellar mass function (SMF). More details are provided in \S~\ref{calibration}.
%Although not used in the calibration, other results of
%the model were visually inspected to ensure their agreement with observations was not seriously compromised by the $z <0.1$ SMF only. Those included 
%$z\approx 0$ atomic and molecular gas scaling relations, the mass-metallicity relation and the $z < 2$ cosmic specific star formation rate density (CSFRD).

Below we briefly summarise the modelling of star formation, stellar feedback, disk instabilities and light emission and 
dust attenuation in the version of \shark\ used in this work, all of which are critical for understanding what drives the predicted UV luminosity function evolution.

\subsection{Star formation modelling}\label{starformation}

The gas in the interstellar medium (ISM) of galaxies is assumed to follow
an exponential profile of half-mass radius  $r_{\rm gas, disk}$ for disks and $r_{\rm gas,bulge}$ for bulges. 
The SFR surface density is calculated assuming a constant molecular gas depletion time,

\begin{equation}
\Sigma_{\rm SFR} = \nu_{\rm SF}\,f_{\rm mol}\,\Sigma_{\rm gas},
\label{eq:SFLaw}
\end{equation}

\noindent where $\nu_{\rm SF}$ is the inverse of the H$_2$ depletion timescale, and
$f_{\rm mol}\equiv \Sigma_{\rm mol}/\Sigma_{\rm gas}$, where $\Sigma_{\rm mol}$ is the molecular gas surface density and
$\Sigma_{\rm gas}$ is the total gas surface density. In this paper we adopt the {\tt BR06} model introduced in 
\citet{Lagos18} (see their \S~4.4.2). The {\tt BR06} model is based on \citet{Blitz06}, 
who found that the H$_2$ to HI ratio, $R_{\rm mol}\equiv \Sigma_{\rm H_2}/\Sigma_{\rm HI}$,
correlates with the local hydrostatic pressure as

\begin{equation}
R_{\rm mol} =  \left( \frac{P}{P_0} \right)^{\alpha_{\rm P}}, 
\label{eq:r_mol}
\end{equation}

\noindent where $P_0$ and
$\alpha_{\rm P}$ are parameters measured in observations of local star-forming galaxies and have values
$P_0/\kappa_{\rm B} = 1,500-40,000\,\rm cm^{-3}\,\rm K$ and $\alpha_{\rm P} \approx 0.7-1$
\citep{Blitz06,Leroy08,Leroy13}. We calculate the hydrostatic pressure from the
surface densities of gas and stars following \citet{Elmegreen89},

\begin{equation}
P = \frac{\pi}{2} \,G\,\Sigma_{\rm gas}\left(\Sigma_{\rm gas} + \frac{\sigma_{\rm gas}}{\sigma_{\star}}\,\Sigma_{\star} \right),
\label{eq:press}
\end{equation}

\noindent where $\Sigma_{\rm gas}$ and $\Sigma_{\star}$ are the
total gas (atomic plus molecular) and stellar surface densities, respectively,
and $\sigma_{\rm gas}$ and $\sigma_{\star}$ are the gas and stellar velocity dispersions, respectively.
The stellar surface density is assumed to follow an exponential profile with
a half-mass stellar radius of $r_{\star,\rm disk}$ for disks and $r_{\star,\rm bulge}$ for bulges.
We adopt $\sigma_{\rm gas}=10\,\rm km\,s^{-1}$ \citep{Leroy08}
 and calculate $\sigma_{\star}=\sqrt{\pi\,G\,h_{\star}\,\Sigma_{\star}}$.
Here, $h_{\star}$ is the stellar scale height,
and we adopt the locally observed relation $h_{\star}=r_{\star}/7.3$ \citep{Kregel02}, with
 $r_{\star}$ being the half-stellar mass radius of the disk or the bulge, depending on the components being 
treated.

Starbursts, which can be triggered by either galaxy mergers or disk instabilities, build up the central bulge of galaxies in \shark.
Thus, when we refer to starbursts, we mean star formation taking place in the central bulge.

There is strong evidence that starbursts follow a similar star formation relation to normal star-forming galaxies
 but with a timescale significantly shorter \citep{Daddi10b,Genzel15,Tacconi18}.
We then adopt the same calculation of $R_{\rm mol}$, $\Sigma_{\rm gas}$ and $P$ above for bulges and disks. 
The implicit assumption is that the gas in both the disk and the bulge is well described by an exponential profile but of different scale radius. 
The only important difference between star formation in disks and starbursts is that we apply a boost factor to the 
star formation efficiency in starbursts $\nu_{\rm SF,\rm burst} = \eta_{\rm burst} \,\nu_{\rm SF}$,
with $\eta_{\rm burst}=15$, broadly consistent with observations \citep{Daddi10b,Scoville16,Tacconi17}. This boost is a critical decision in the resulting UV LF, as we will discuss in \S~\ref{uvlfsec}.

We note that the star formation model above is informed by observations primarily at $z<2$ and is assumed to 
apply to all cosmic times. 

\subsection{Stellar feedback}

\shark\ separates stellar feedback into two main components: the outflow rate of the gas that escapes from the galaxy,
$\dot{m}_{\rm outflow}$, and the ejection rate of the gas that escapes from the halo, $\dot{m}_{\rm ejected}$.

We define $\dot{m}_{\rm outflow}=\psi\,{\rm f}(z,V_{\rm circ})$, where $\psi$ is the instantaneous SFR, $z$ is the redshift and $V_{\rm circ}$
is the maximum circular velocity of the galaxy. The ejection rate of the halo is $>0$
only in the case where the injected total energy of the outflow is larger than the specific binding energy
of the halo. 

\citet{Muratov15} used the FIRE simulation suite to estimate several properties of the stellar driven outflows,
including the terminal wind velocity, $V_{\rm w}$. \citet{Muratov15} found that

\begin{equation}
\frac{V_{\rm w}}{\rm km\,s^{-1}}=1.9\,\left(\frac{V_{\rm circ}}{\rm km\,s^{-1}}\right)^{1.1}.
\end{equation}

\noindent We use this terminal velocity to compute the excess energy that is used to eject gas out of the halo as
\begin{equation}
{\rm E}_{\rm excess} = \epsilon_{\rm halo}\,\frac{V^2_{\rm w}}{2} \psi\,{\rm f}(z,V_{\rm circ}).
\label{eq:epshalo}
\end{equation}

\noindent Here $\epsilon_{\rm halo}=10$ in the model we use in this paper, informed by the EAGLE cosmological hydrodynamical simulations \citep{Mitchell20}. The net ejection rate is calculated as,

\begin{equation}
\dot{m}_{\rm ejected} = \frac{{\rm E}_{\rm excess}}{V^2_{\rm circ}/2} - \dot{m}_{\rm outflow}.
\end{equation}

\noindent If $\dot{m}_{\rm ejected} < 0$ no ejection from the halo takes place and we limit
$\dot{m}_{\rm outflow}$ to ${\rm E}_{\rm excess}/(V^2_{\rm circ}/2)$.

\begin{figure}
\begin{center}
\includegraphics[trim=2mm 5mm 2mm 2mm, clip,width=0.5\textwidth]{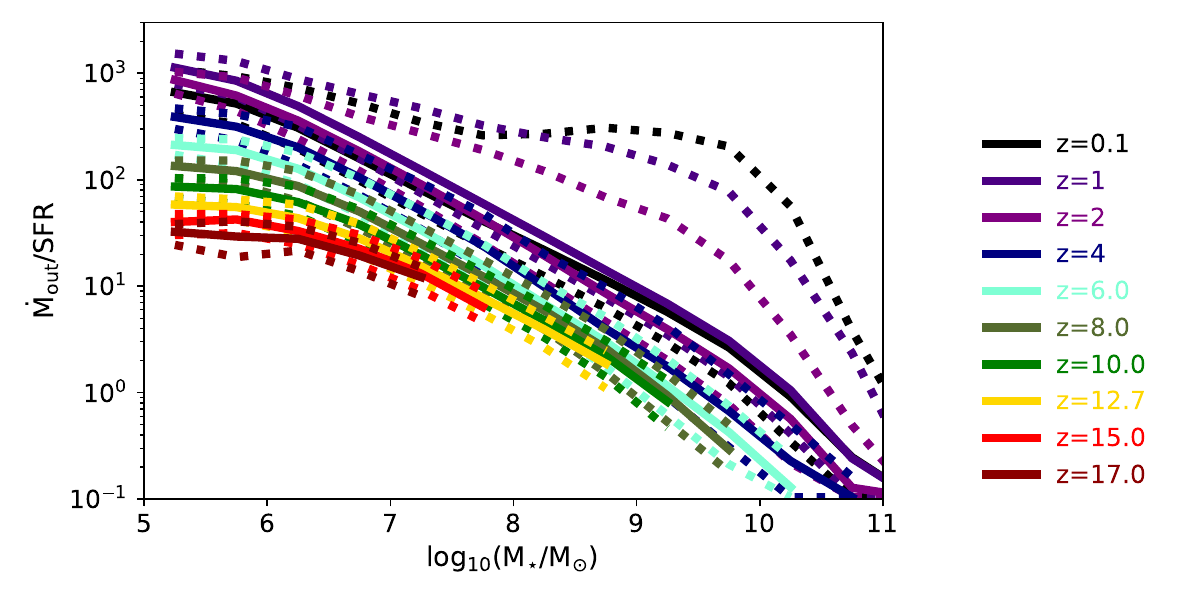}
\caption{The mass loading of winds driven by stellar feedback alone as a function of stellar mass for all galaxies in \shark\ at redshifts $z=0.1$ to $z=17$, as labelled. Solid lines show the medians, and dashed lines show the $16^{\rm th}-84^{\rm th}$ percentile ranges.} 
\label{mass_loading}
\end{center}
\end{figure}

In this paper, we adopt the {\tt Lagos13} model introduced in \S~4.4.3 in \citet{Lagos18}, which is based on an analytic model of supernovae-inflated bubbles in a multi-phase ISM introduced in \citet{Lagos13}. In that model, star formation was assumed to be distributed in clouds along the disk, so that the radial profile of the SFR follows the molecular gas surface density profile using the \citet{Leroy08} molecular gas conversion efficiency. The inflated bubbles go through three phases, from an early adiabatic growth, a snowplough-phase, finishing in a momentum-driven expansion phase. The outflow rate is defined as the mass of the bubbles that break out from the ISM. \citet{Lagos13} used that model coupled to the semi-analytic model GALFORM \citep{Lagos12}, to parametrise the outflow rate based on the circular velocity of the halo and redshift, which is the equation that was implemented in \citet{Lagos18}. In \shark\ v2.0 
we adopted some small modifications to the model presented in \citet{Lagos18} following the equations below:

\begin{eqnarray}
{\rm f} &=& \rm max\left[\epsilon_{\rm disk}\, \left(\frac{V_{\rm circ}}{v^{\prime}_{\rm hot}}\right)^{\beta}, 0.1\right],\label{SNLAG13}\\
v^{\prime}_{\rm hot}  &=& v_{\rm hot} \, t_{\rm age}^{\rm z_{\rm P}},\label{vel_power}
\end{eqnarray}

\noindent where $t_{\rm age}$ is the age of the universe, and 
$v_{\rm hot} = 120\,\rm km\,s^{-1}$, $\beta = 3.79$, $\rm z_{\rm P} = 0.13$, $\epsilon_{\rm disk}=1$. Note that the choice of parameters leads to mass loading values that are inversely correlated with stellar mass, and that become smaller with increasing redshift. This is shown in Fig.~\ref{mass_loading}. \shark\ predicts a strong evolution of the mass loading ($\equiv \dot{m}_{\rm outflow}/\rm SFR$) especially for low-mass galaxies, $M_{\star}<10^7\,\rm M_{\odot}$, with values ranging from $\approx 30$ at $z=17$ to $\approx 10^2-10^3$ at $z=0.1$. At the higher mass end, $M_{\star}>10^8\,\rm M_{\odot}$, the scatter evolves strongly driven by the flattening of the stellar-halo mass relation above a halo mass scale of $\approx 10^{12}\,\rm M_{\odot}$. We highlight that although stellar feedback is indeed weaker at $z\gtrsim 10$ in our fiducial \shark\ model, the mass loading is still very significant in the dwarf galaxy regime, and thus inconsistent with the idea of feedback-free starbursts, which has been discussed as a potential necessity to reproduce the high-z UV LF \citep{Dekel23}. 
%implies that stellar feedback is monotonically stronger with increasing redshift at fixed $V_{\rm circ}$. ADD A FIGURE TO SHOW THE EVOLVING 
%MASS LOADING MAYBE WITH STELLAR MASS TO MAKE IT MORE TRACTABLE FOR OBSERVERS. 

\shark\ also implements a model for reincorporation of the gas ejected by stellar feedback from halos, but the timescale of reincorporation is irrelevant 
at $z>6$, which is the focus of this paper. We refer the reader to \S~4.4.5 in \citet{Lagos18} for details on the modelling of reincorporation.

\subsection{Triggering of starbursts}

Starbursts can be triggered in two different ways, as described below. Regardless of the triggering mechanism, the amount of stars formed  
is calculated as described in \S~\ref{starformation}.\\

\subsubsection{Galaxy mergers}

When DM halos merge, we assume that the galaxy hosted by the main progenitor halo becomes
the central galaxy, while all the other galaxies become satellites orbiting
the central galaxy. These orbits gradually decay towards the centre due to energy and angular
momentum losses driven by dynamical friction with the halo material, including other satellites.
When the DM subhalo hosting a satellite galaxy disappears from the halo catalogues (most likely 
due to dropping below the minimum number of particles required for a ``detection''), those satellite galaxies 
become ``orphans''.  
We calculate a dynamical friction timescale following \citet{Poulton2021} for these orphans, and merge those with the central galaxy 
once that clock goes to zero.

Depending on the amount of gas and baryonic mass involved in the galaxy merger, a starburst 
can be triggered. We define the total mass of gas plus stars of the primary (most massive) and secondary (least massive) galaxies involved in a merger as:  
$M_{\rm p}=M_{\rm cold,p}+M_{\star,\rm p}$ and
$M_{\rm s}=M_{\rm cold,s}+M_{\star,\rm s}$, respectively. We define the galaxy mass ratio as $M_{\rm s}/M_{\rm p}$, and
the gas fraction of the primary galaxy as $M_{\rm cold,p}/M_{\rm p}$. The latter is strictly $\le 1$. The outcome of the merger follows the list below:

\begin{itemize}
\item If $M_{\rm s}/M_{\rm p}>0.25$, there is a major merger. In this case, all the stars
present are rearranged into a spheroid. In addition, any cold gas in the merging system is assumed to undergo
a starburst and the stars formed are added to the spheroid component. 
\item If $0.1<M_{\rm s}/M_{\rm p}\le 0.25$, there is a minor merger. In this case all the stars
in the secondary galaxy are accreted onto the primary galaxy's spheroid, leaving intact the stellar disk
of the primary. In minor mergers the triggering of a starburst depends on the cold gas content of the
primary galaxy. If $M_{\rm cold,p}/M_{\rm p}>0.3$, a starburst 
is triggered as the perturbations introduced by the secondary galaxy should suffice to drive all the cold gas from
both galaxies on to the new spheroid. If $M_{\rm cold,p}/M_{\rm p}<0.3$,
the gas mass of the secondary is accreted by the disk of the primary.
\item If $M_{\rm s}/M_{\rm p}\le 0.1$, the primary disk is left unperturbed. As before, the stars accreted
from the secondary galaxy are added to the spheroid, but the overall gas component (from both galaxies) stays in the disk,
along with the stellar disk of the primary.
\end{itemize}

\subsubsection{Disk Instabilities}
If the disk becomes sufficiently massive that its self-gravity is dominant, then it is unstable to small perturbations
by minor satellites or dark matter substructures. The criterion for instability was described in \citet{Ostriker73} and \citet{Efstathiou82} as,

\begin{equation}
\epsilon=\frac{V_{\rm circ}}{\sqrt{1.68\,G\, M_{\rm disk}/r_{\rm disk}}}.
\label{DisKins}
\end{equation}

\noindent Here, 
$r_{\rm disk}$ is the half-mass disk radius and
$M_{\rm disk}$ is the disk mass (gas plus stars). The numerical factor $1.68$ converts the disk half-mass radius into a scalelength, assuming an exponential profile.
If $\epsilon<0.98$ the disk is considered to be unstable.
In \shark, gas and stellar disks can have different sizes, and thus to evaluate Eq.~\ref{DisKins} we compute a mass-weighted $r_{\rm disk}$ between the two disk
components. If the disk is unstable, stars and gas in the disk are accreted onto the spheroid and the radial gas inflow drives a starburst.
Several detailed hydrodynamical simulations have been performed to study the effect of these ``violent disk instabilities''
 and have shown that they can form galaxies with steep stellar profiles, 
similar to early-type galaxies (e.g. \citealt{Ceverino15}, \citealt{Zolotov15}).
 
\subsection{Light emission and dust attenuation in galaxies}\label{dustsec}

The modelling of light emission and dust attenuation was introduced in \citet{Lagos19}. Here we provide a brief overview. 

We calculate the light emitted by galaxies using {\sc ProSpect} \citep{Robotham20}, which combines the GALEXEV stellar synthesis libraries 
\citep{Bruzual03} with the two-component dust attenuation model of \citet{Charlot00}. {\sc ProSpect} also calculates nebular emission lines 
using {\sc Mappings} \citep{Levesque10} and assuming that the ionisation parameter scales with the gas metallicity, 
following \citet{Orsi14}. In {\sc ProSpect}, the default is to assume an escape fraction of 0, unless explicitly set by the user--thus, our assumption here is 0.  
%(XXX say something about the escape fraction assumed for the emission lines). 

{\sc ProSpect} takes as input  
the star formation and chemical enrichment history of each galaxy component (disk and bulge), further separating the bulge component by the stars 
that were either accreted or formed via a starburst triggered by galaxy mergers or disk instabilities. This means that the emission of any one galaxy can 
be split into three contributing components: a disk, the part of the bulge coming from galaxy mergers, and the part coming from disk instabilities.

After the emitted light is calculated, we attenuate the light using a two-component dust model, diffuse dust and birth clouds (with the disk and the bulge having distinct dust components). 
We first attenuate and re-emit the light due to birth clouds, and then attenuate and re-emit the light due to the diffuse dust. 
The \citet{Charlot00}, hereafter CF00, functional form for the attenuation curves for stars in the diffuse ISM and birth clouds, which are written as follows: 

\begin{eqnarray}
\tau_{\rm ISM} &=& \hat{\tau}_{\rm ISM}\, (\lambda/5500\text{\AA})^{\eta_{\rm ISM}}
\label{taus_screen},\\
\tau_{\rm BC} &=& \hat{\tau}_{\rm BC}\, (\lambda/5500\text{\AA})^{\eta_{\rm BC}},\label{tau_bc}
\end{eqnarray}

\noindent respectively, where $\hat{\tau}_{\rm ISM}$ and $\hat{\tau}_{\rm BC}$ are the optical depth at $5500$\AA\ in the diffuse ISM and birth clouds, respectively, and $\eta_{\rm ISM}$ and $\eta_{\rm BC}$ are the power-law indices that control the dependence on wavelength for the diffuse ISM and birth clouds, respectively. In \shark, we scale the
CF00 parameters above depending on local properties of galaxies.

In the case of birth clouds, we compute the optical depth from the
gas metallicity and a typical cloud surface density (see equation~$6$ in
\citealt{Lagos19}). 
The cloud surface density is defined as $\Sigma_{\rm gas,cl}={\rm max}[\Sigma_{\rm MW,cl},\Sigma_{\rm gas}]$, with $\Sigma_{\rm MW,cl}=85\,\rm M_{\odot}\,pc^{-2}$ \citep{Krumholz09}, $\Sigma_{\rm gas}$ being the diffuse medium gas surface density, $0.5\,\Sigma_{\rm gas}/\pi\,r^2_{\rm 50}$, and 
$r_{\rm 50}$ being the half-gas mass radius. The wavelength power-law index is fixed to $\eta_{\rm BC}=-0.7$, following the inference of \citet{Charlot00}. 

For the diffuse dust, we use the scaling derived from the radiative transfer 
analysis of EAGLE galaxies at $0\le z \le 2$ using SKIRT \citep{Camps15} proposed by \citet{Trayford20}. 
 These consist of $\hat{\tau}_{\rm ISM}$ and $\eta_{\rm ISM}$ varying as a function of the dust surface density, $\Sigma_{\rm dust}$. These relations were shown to be redshift independent at least up to $z=2$ by \citet{Trayford20}. We compute $\Sigma_{\rm dust}$ independently for the disks and bulges of galaxies. To compute $\Sigma_{\rm dust}$, we include the effect of inclination as described in \citet{Lagos19} (see their equation~2).

\begin{figure}
\begin{center}
\includegraphics[trim=2mm 4mm 2mm 2mm, clip,width=0.5\textwidth]{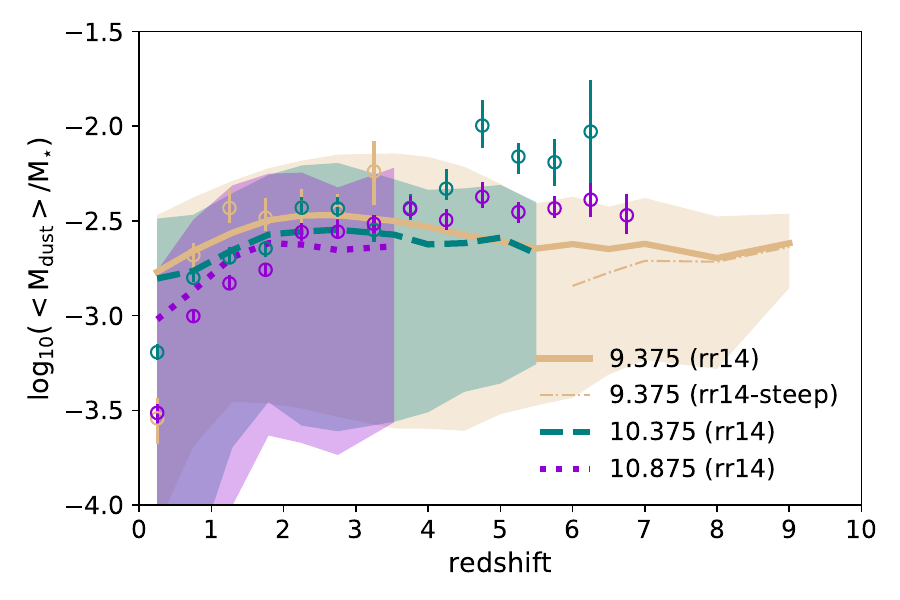}
\caption{Mean dust-to-stellar mass ratio evolution for galaxies in three bins of stellar mass (of width $0.25$~dex) in \shark\ from $z=0$ to $z=7$. Lines show the means and the shaded regions show the $16^{\rm th}-84^{\rm th}$ percentile ranges. The labels show the mid point of the adopted stellar mass bins. Only bins and redshifts with $\ge 20$ galaxies are shown. For clarity, we only show results from \shark\ combined with L800 (see \S~\ref{simsused}). The three stellar mass bins over the whole redshift range are shown for \shark\ adopting the ``rr14'' scaling (Eq.~\ref{eq.mdust}), while for the ``rr14-steep'' we only show the mean in lowest mass bin and only at $z\ge 6$ to demonstrate it converges well with ``rr14'' in the regime of interest for the UVLF in this work. The medians for the two higher mass bins in the ``rr14-steep'' scaling (Eq.~\ref{eq.mdust2}) are almost exactly the same as the results obtained with the ``rr14'' scaling, so we do not show them here. Observational results from a stacking analysis in  \citet{Casey26} are shown as symbols using the same colours as the \shark\ results.} 
\label{mdustmstar}
\end{center}
\end{figure}
 
Dust masses in \shark\ are calculated for each galaxy component
from their gas mass and metallicity, following the best-fitting relation
between the dust mass and the latter two galaxy properties in the local
Universe of \citet{Remy-Ruyer14}. We consider two fits presented in \citet{Remy-Ruyer14}, the best fit, which takes the shape:

\begin{eqnarray}
    {\rm log}_{10}(D/M) = 2.21 - {\rm log}_{10}(Z/Z_{\odot})\,\,\,\,\,\,\, {\rm if}\, {\rm log}_{10}(Z/Z_{\odot})>-0.59\,\,\,\label{eq.mdust}\\
   {\rm log}_{10}(D/M) = 0.96 - 3.1\, {\rm log}_{10}(Z/Z_{\odot})\,  {\rm if}\,{\rm log}_{10}(Z/Z_{\odot})\le -0.59\nonumber
\end{eqnarray}
    
\noindent which we refer to as ``rr14''; and a steeper relation is assumed with a break at higher gas metallicities, which is favoured by the more recent observations of \citet{DeVis19}, which we refer to as ``rr14-steep'' and takes the shape:

\begin{eqnarray}
    {\rm log}_{10}(D/M) = 2.21 - {\rm log}_{10}(Z/Z_{\odot})\,\,\,\,\,\,\,\,\,\, {\rm if}\, {\rm log}_{10}(Z/Z_{\odot})>-0.16\,\,\,\label{eq.mdust2}\\
   {\rm log}_{10}(D/M) = 1.66 - 4.43\, {\rm log}_{10}(Z/Z_{\odot})\,  {\rm if}\,{\rm log}_{10}(Z/Z_{\odot})\le -0.16\nonumber
\end{eqnarray}
%MAKE A PLOT COMPARING THE DUST/METAL RATIO VS METALLICITY 
%THAT WE USE AND EARLY UNIVERSE GALAXIES TO SEE HOW WELL IT APPLIES.

Recently, \citet{Casey26} carried out one of the largest {\it JWST} stacking campaigns to derive average dust masses for galaxies of stellar masses $>10^9\,\rm M_{\odot}$ at $0\le z\le 7$. We compare the derived mean dust-to-stellar mass ratio evolution in \shark\ with the results presented in \citet{Casey26} in Fig.~\ref{mdustmstar} for three stellar mass bins. Here we show results of \shark\ using the ``rr14'' scaling of Eq.~\ref{eq.mdust}. Overall, the agreement with the observations is quite good, particularly at $0.5\lesssim z\lesssim 4$, which is the range at which the sample in \citet{Casey26} has the best statistics (see their fig.~1). At higher redshifts, \shark\ does not produce enough massive galaxies to calculate the mean and percentile ranges. \shark\ reproduces quite well the trend of massive galaxies showing the strongest evolution in the dust-to-stellar mass ratio at $z\lesssim 2$, and the plateau in the mass ratio seen for most stellar mass bins at higher redshifts. This result  gives us confidence that the dust reservoirs we use to compute the dust attenuation of galaxies in \shark\ is reasonable in the regime in which observations currently exist.

\begin{table*}
        \setlength\tabcolsep{2pt}
        \centering\footnotesize
        \caption{Specifications of the $N$-body simulations used in this work. For micro-SURFS and medi-SURFS, the outputs are composed of $200$ snapshots from $z=30$ to $z=0$, while for Hi-SURFS, the simulation contains $50$ snapshots between $z=30$ and $z=5$. L800 is the only simulation we use that does not form part of the SURFS suite. It contains 271 snapshots from $z=127$ to $z=0$. Here, cMpc and ckpc refer to comoving Mpc and kpc, respectively. }
        \begin{tabular}{@{\extracolsep{\fill}}l|ccccccc|p{0.45\textwidth}}
                \hline
                \hline
            Name & Box size & Number of & Particle Mass & Softening Length & Minimum $M_{\rm halo}$ & output redshift range & \# snapshots $z\ge 6$\\
        & $L_{\rm box}$ [${\rm cMpc}/h$] & Particles $N_{\rm p} $ & $m_{\rm p}$ [${\rm  M}_{\odot}/ h$] & $\epsilon$ [${\rm ckpc}/h$] & $M_{\rm min}\,[{\rm M_{\odot}}/h]$\\
        \hline
    L40N512  (micro-SURFS)   & $40$  & $512^3$   & $4.13\times10^7$ & 2.6 &$8.6\times 10^8$ & $30-0$ & 89\\
    L50N2048 (Hi-SURFS) & $50$ & $2048^3$ & $1.26\times10^6$& 0.98 & $2.52 \times 10^7$&$30-5$ & 46\\
    L210N1536 (medi-SURFS)  & $210$ & $1536^3$  & $2.21\times10^8$ & 4.5  & $4.42\times 10^9$ & $30-0$ & 89\\
    L542N5040 (L800) & $542.26$ & $5040^3$ & $1.06\times 10^8$ & 2.3 & $2.12\times 10^9$& $127-0$ & 82\\ 
    \hline
        \end{tabular}
        \label{tab:sims}
\end{table*}

If we instead adopt the ``rr14-steep'' scaling of Eq.~\ref{eq.mdust2}, we find the results for the two massive bins in Fig.~\ref{mdustmstar} do not change, as those galaxies tend to have gas metallicities $\rm log_{10}(Z/Z_{\odot})\gtrsim -0.16$. For the lower mass bin, however, we find lower dust-to-stellar mass ratio with the ``rr14-steep'' scaling at $z\lesssim 6$ than what is obtained for the ``rr14'' scaling, but the medians of the two models tend to produce convergent dust-to-stellar mass ratios at $z\gtrsim 6$, as shown by the thin dot-dashed line in Fig.~\ref{mdustmstar}. Appendix~\ref{dustatt} shows the impact that the choice of dust-to-metal ratio scaling has on the predicted attenuated UVLF at $z\ge 6$, showing that the ``rr14-steep'' scaling tends to produce a brighter end of the UVLF, especially at $z\gtrsim 13$. In what follows, we show the predicted UVLF using the ``rr14-steep'' scaling, but also the unattenuated UVLF to test whether \shark\ produces enough UV bright galaxies at cosmic dawn compared with the latest observations.

%The authors compared their results with \shark\ predictions, using the scales of Eq.~\ref{eq.mdust} in their fig.~20 showing excellent agreement. The default choice throughout this paper is ``rr14'',  unless otherwise stated. 

In this work, we focus on the emission of galaxies at a rest-frame wavelength of $1,500$\AA, which we measure as the rest-frame AB absolute magnitude of galaxies using a top-hat filter of width $100$\AA\ around the target wavelength. \citet{Lagos19} tested the predicted UVLF in \shark\ at $z=3-10$ using the rest frame FUV GALEX filter vs the top-hat filter adopted here, finding negligible differences between them. Thus, we consider the choice of band here appropriate to study the UVLF at $z\ge 6$. 

Another important aspect to highlight is that for young stellar populations,\citet{Robotham25} show that in predicting/inferring the flux at around a rest-frame of $1,500\AA$ the choice of stellar population library matters. Adopting BPASS \citep{Stanway18}, a common choice in high-z studies, produces approximately a factor of $\approx 2$ higher fluxes relative to most of the other commonly adopted libraries (see to-left panel in fig.~14 of \citealt{Robotham25}), including GALEXV adopted here. This is the case at fixed metallicity, age and element abundances.
\citet{Robotham25} show that differences at the FUV between different libraries are largest when very young (OB stars) stars are dominant (i.e. young populations), and when stars are very old (multi Gyr UV upturn galaxies) where PNe and white dwarfs can significant impact UV predictions.
 The difference among UV predictions across stellar population libraries is an underlying systematic effect impacting derived UV luminosities both in simulations and observations that needs to be considered when comparing them. 
%a broad-band shape in the rest-frame consistent with the GALEX FUV band does not noticeably impact our predicted UVLF. 

\subsection{$N$-body simulations used in this work}\label{simsused}

Here we use a suite of $N$-body simulations, including three of the \surfs\ suite \citep{Elahi18}, 
%
%version of \shark\ running over 
and the P-Millennium $N$-body simulation (hereafter L800). The idea of employing a suite of $N$-body simulations, instead of a single one as done more commonly in the literature, is to assess the impact of resolution on the predictions. We will be focusing on resolution as defined by the particle mass employed but also the time cadence of the snapshots. This allows us to also have runs of varying cosmological volumes. 
%and the XXXX. 

The \surfs\ simulations adopt a $\Lambda$CDM \citep{Planck15} cosmology. The cosmological parameters correspond to a total matter, baryon and $\Lambda$ densities of $\Omega_{\rm m}=0.3121$, $\Omega_{\rm b}=0.0491$ and $\Omega_{\rm  L}=0.6879$, respectively, with a Hubble parameter of $H_{\rm 0}=h \, 100\,\rm Mpc\, km\,s^{-1}$ with $h=0.6751$, scalar spectral index of $n_{\rm s}=0.9653$ and a power spectrum normalization of $\sigma_{\rm 8}=0.8150$. Compared to the published suite in \citet{Elahi18}, the runs used here have been fully re-processed using the {\sc SWIFT} code \citep{Schaller24}, on the Setonix supercomputer at the Pawsey Supercomputing Centre. In this paper, we use the L210N1536 (medi-SURFS), L040N0512 (micro-SURFS) and L50N2048 (Hi-SURFS) simulations, with the specifications shown in   Table~\ref{tab:sims}. medi-SURFS and micro-SURFS contain $200$ snapshots; at $z\ge 6$ snapshots typically have a time span between them of $\approx 2.6-24$~Myr. %These two runs contain 88 snapshots between $z=6$ and $z=30$. 

Hi-\surfs\ is a run specifically produced for early universe studies, and has been run down to $z=5$, using $50$ snapshots, typically having a time span between snapshots of $\approx 5-43$~Myr at $z\ge 6$. Note that the time cadence of this run is twice poorer than the other \surfs\ runs above. We will come back to the impact of this choice later. 

Halo and subhalo catalogues and the merger trees for the \surfs\ suite were produced using {\sc HBT-Herons} \citep{Forouhar-Moreno25}. \citet{Chandro-Gomez25a} demonstrated {\sc HBT-Herons} produces very clean merger trees and halo catalogues that largely alleviate typical numerical artefacts which are common in other codes. Those include central/satellite subhalo flag swapping, and incorrect identification of progenitors, which lead to the sudden appearance of massive halos at late cosmic times. These problems are more prevalent at low redshift and in massive halos ($\gtrsim 10^{12}\,\rm M_{\odot}$), but are minimal in the early universe. \citet{Chandro-Gomez25a} introduced a methodology to avoid these numerical artefacts at the SAM level, which we adopt in the runs presented here (see Appendix~\ref{newcalibration} for more details). 

L800 was introduced in \citet{Baugh19} and the cosmological parameters correspond to a total matter, baryon and $\Lambda$ densities of $\Omega_{\rm m}=0.307$, $\Omega_{\rm b}=0.04825$ and $\Omega_{\Lambda}=0.693$, respectively, with a Hubble parameter of $H_{\rm 0}=h \, 100\,\rm Mpc\, km\,s^{-1}$ with $h=0.6777$, scalar spectral index of $n_{\rm s}=0.9611$ and a power spectrum normalisation of $\sigma_{\rm 8}=0.8288$. L800's volume and particle mass are  presented in Table~\ref{tab:sims}.  L800 has $271$ snapshots, typically having a time cadence between snapshots of
$\approx 4-26$~Myr at $z\ge 6$, similar to micro-SURFS and medi-SURFS. 
Halo catalogues for L800 were constructed using {\sc Subfind} \citep{Springel01} and merger trees were built using D-Trees+DHalo \citep{Jiang14}. 
%Haloes with $\ge 20$ particles are included in the catalogues, giving a minimum halo mass of $2.12\times 10^9\,\rm M_{\odot}\,h^{-1}$. 
\citet{Chandro-Gomez25a} showed that the merger trees in L800 had frequent numerical artefacts, but those are minimised or entirely avoided by introducing the fixing methods suggested by \citet{Chandro-Gomez25a}, which here we adopted. 

\subsection{Model calibration}\label{calibration}

The fiducial model adopted here was calibrated to the $z=0.1$ SMF of \citet{Li09}, as described in \citet{Lagos24}, using medi-SURFS, and the automatic Particle Swarm Optimisation (PSO) package, {\tt optim}, that is part of \shark.  

Although we only use the $z=0.1$ SMF as a numerical constraint for {\tt optim}, we visually inspected other results of the model to ensure we were not seriously compromising the agreement with observations by using the $z=0$ SMF only. For example, we inspect the atomic and molecular gas scaling relations at $z=0$, the mass-metallicity relation at $z=0$, and the cosmic star formation rate density (CSFRD)  at $z<2$ and make sure that they look reasonable. This does not mean that we change the parameters from the best-fitting ones but instead we used the visual inspection of other results to draw reasonable prior ranges in some of the parameters that we input to {\tt optim}. 

Once a good calibration with medi-SURFS was reached with the $z=0.1$ SMF, we simply adopted the exact same parameters for the runs produced with the other simulations in Table~\ref{tab:sims}. In Appendix~\ref{newcalibration}, we introduce the parameters that are different from those adopted in \citet{Lagos24}.  

\section{The UV luminosity function at $z\ge 6$}\label{uvlfsec}

\begin{figure*}
\begin{center}
\includegraphics[trim=23mm 8mm 2mm 23mm, clip,width=0.99\textwidth]{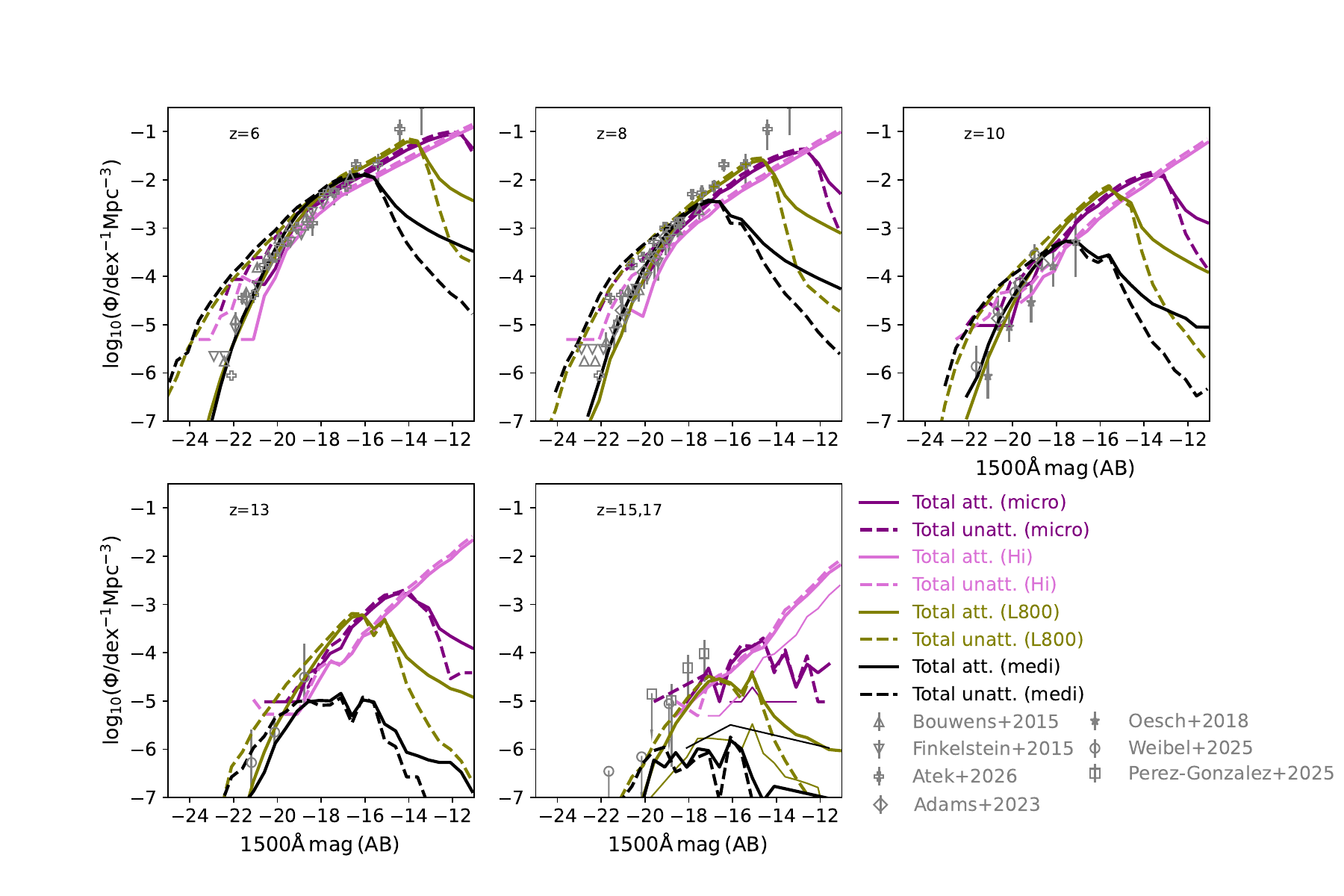}
\caption{The rest-frame UV luminosity function (UVLF) from $z=6$ to $z=17$ in \shark, as labelled in each panel. We show the unattenuated and attenuated UVLF predicted by the same model run over medi-SURFS, micro-SURFS and L800. For the panel at $z=15,17$, we show with thick lines $z=15$ predictions, and with thin lines, $z=17$ predictions. We opt to do this in this panel only due to (i) the wide redshift range mapped by the drop-out technique used in observations, and (ii) the extremely rapid evolution of the UV LF at these very early epochs. 
Observational constraints from \citep{Bouwens15,Finkelstein15,Oesch18,Adams23,Weibel25b,Perez-Gonzalez25,Atek26} are shown with symbols, as labelled. \shark\ predicts a UVLF that agrees very well with observations once all boxes are combined. In general, the large volume (lower resolution) boxes are required to reproduce the bright end of the UVLF, while the smaller (higher resolution) boxes are required to reproduce the faint end of the UVLF, highlighting the importance of resolution to push towards the low-mass halos that host galaxies with UV magnitudes $<-15$.} 
\label{UVLF}
\end{center}
\end{figure*}

\begin{figure}
\begin{center}
\includegraphics[trim=2mm 2mm 2mm 2mm, clip,width=0.49\textwidth]{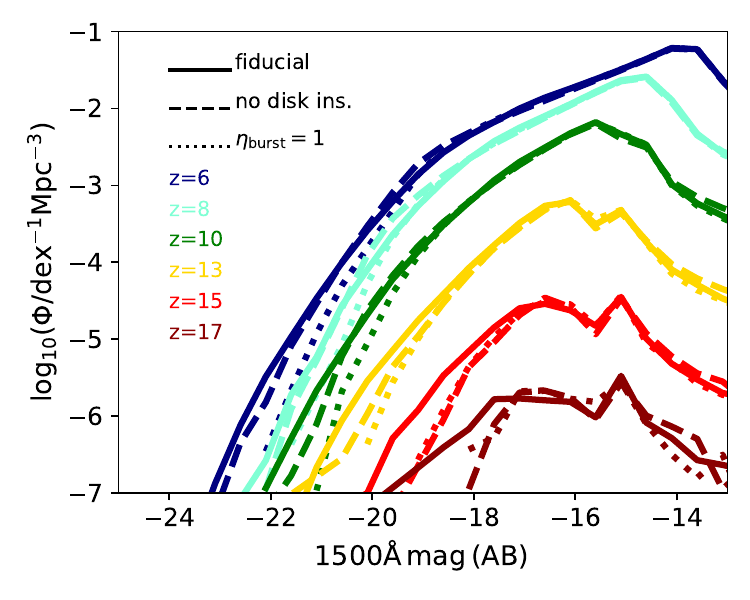}
\caption{The rest-frame unattenuated UV luminosity function (UVLF) at $z=6,\, 8,\,10,\,13,\,15,\,17$, as labelled, in \shark\ when run over the L800 box. We show three realisations of the model, our fiducial run (solid lines), a run with identical parameters except that disk instabilities have been turned off (dashed lines), and a run in which we adopt $\eta_{\rm burst}=1$, making the molecular gas surface density star formation efficiency universal (dotted lines).} 
\label{UVLFnodiskins}
\end{center}
\end{figure}

\begin{figure*}
\begin{center}
\includegraphics[trim=22mm 8mm 29mm 23mm, clip,width=0.99\textwidth]{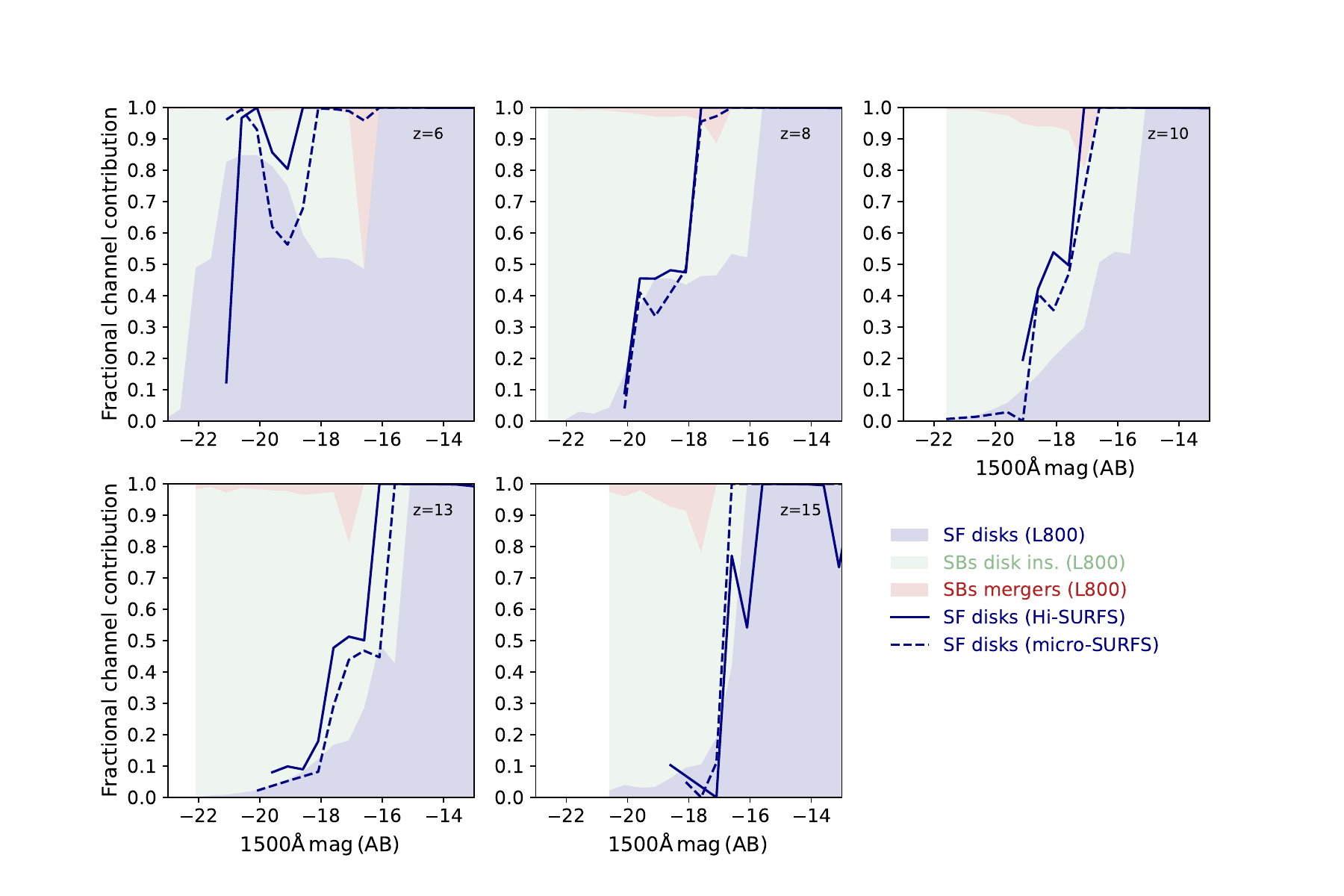}
\caption{Fractional contribution to the UV luminosity by the three star formation channels in \shark\ v2.0. Here we show the \shark\ run using L800 with shaded regions. Blue, green and red shaded regions show the contribution from star formation in disks, and starbursts associated with disk instabilities and mergers, respectively. We also show with solid and dashed lines, the corresponding contribution from star formation in disks in \shark\ when run using Hi-SURFS and micro-SURFS, both are smaller boxes than L800 but have much higher resolution (see Table~\ref{tab:sims}). The fractional contribution values 
%
%and the values 
are based on the median contribution in bins of UV magnitude. Starbursts triggered by disk instabilities are a major source of luminosity in UV bright galaxies, especially at $z\gtrsim 8$. On the other hand, the stochastic nature of galaxy mergers makes them a relatively small contributor across the UV magnitude range.}
%As in Fig.~\ref{UVLF} but showing the contribution to the total attenuated UVLF from  from starbursts triggered by disk instabilities (red lines), by galaxy mergers (green lines) and  star formation in galaxy disks (blue lines). \shark\ predicts that at all redshifts, the bright-end of the UVLF is dominated by starbursts which have been triggered by disk instabilities.} 
\label{UVLFchannels}
\end{center}
\end{figure*}

Fig.~\ref{UVLF} shows the UVLF predicted by \shark\ at $6\le z\le 17$, when run over all the simulation boxes of Table~\ref{tab:sims}. We show intrinsic and attenuated UVLF, noting that at $z\ge 13$, the attenuation of dust in the UV emission of galaxies is negligible. However, even at $z=10$, the bright end of the UV LF, $M_{\rm UV}\lesssim -21$ is highly affected by dust attenuation, with the attenuated and unattenuated UV LF differing by up to $1.5$~mag.

All the observational measurements shown in Fig.~\ref{UVLF} correspond to inferences based on the drop-out technique and photometric redshifts, which means that any one set of observations corresponds to a redshift range rather than a specific cosmic time. This distinction is important as the UVLF is evolving remarkably fast in the first few hundred Myrs. For instance, the measurements of \citet{Perez-Gonzalez25} and \citet{Weibel25b} correspond to a redshift range of $\approx 15-19$ and $z \approx  15.3 - 21.9$, respectively, with the median being at $z\approx 17$. Thus, we show in the highest redshift panel the UVLF predicted by \shark\ at $z=15$ and $17$. We also note that we show the measurements of \citet{Atek26} at $z=6$ and $z=8$, as the redshift range probed by those observations is $z\approx 6-8$, with a median redshift of $7$.   

We find remarkable agreement with the observations across the redshift range probed within the uncertainties reported. Given the large redshift range probed by the observations we show in the highest redshift panel ($z\approx 15-20$), it is not surprising that the best agreement with \shark\ is obtained at $z=15$. Similarly, at the faint end of the UVLF and at $z\le 8$, 
%It may not be surprising then that 
the best agreement between our predictions and the observations of \citet{Atek26} is obtained at $z=6$--this is expected given the fast evolution of the faint end of the UVLF between $z=8$ and $6$. We also remind the reader that most of these derived UVLF in observations assume the BPASS stellar population library, which for young stellar populations produces $\approx 2$ times the flux that GALEXV (the adopted library in \shark) produces at the rest-frame $1,500\AA$. This means that differences of a factor $\lesssim 2$ are well within that systematic effect (see discussion in \S~\ref{dustsec}).

It is worth noting the significant impact of resolution on the range of UV magnitudes sampled by each run. The \shark\ run using medi-SURFS struggles to produce galaxies at $z>13$ because it is sampling higher mass halos preferentially compared to micro-SURFS, Hi-SURFS and L800. It is also clear that to push down to $M_{\rm UV}=-13$, it is necessary to resolve halos of masses $10^8\,\rm M_{\odot}$.

Another interesting feature is the large differences between runs with modestly different resolutions. For example, L800 has a mass resolution $\approx 2.5$ times better than medi-SURFS, but that translates into a coverage of the UVLF more than 2 magnitudes wider in L800 compared with medi-SURFS. This is clear from the number densities in L800 starting to drop at around $M_{\rm UV} \approx -15$ compared with $M_{\rm UV}\approx -18$ in medi-SURFS at $z=13$. This large difference stems from the steepness of the galaxy mass-halo mass relation in the dwarf-galaxy regime, which in \shark\ is similar to the \citet{Moster13} stellar-halo mass relation slope at halo masses $\lesssim 10^{12}\,\rm M_{\odot}$. We come back to the UV-halo mass scaling later when discussing Fig.~\ref{UVscaling}.  
%At halo masses $\gtrsim 10^{11}\,\rm M_{\odot}$ the relation ceases to evolve with redshift, with the medians staying consistent across $6\le z\le 10$. 

We also highlight the difference between Hi-SURFS and the other simulations. The \shark\ run using Hi-SURFS predicts a UVLF offset in number density relative to micro-SURFS and L800, with the difference becoming smaller as we go from $z\approx 13$ to $z\approx 6$. This difference is resolution-driven. Hi-SURFS's outputs have a time cadence that is approximately twice poorer than the other simulations, which directly affects how finely we can track the early growth of halos and thus, the impact that has on galaxy evolution. From the runs we have tested, we find that in order to have convergence in the predicted UV LF, one needs $\gtrsim 70$ snapshots at $z>6$. This statement has to be taken with caution, however, as it results from our tests running \shark\ on a variety of $N$-body runs, and moreover, the particular \shark\ version we use in this work. It may well be the case that convergence against the time cadence of the $N$-body simulation employed is model-dependent. It would not be surprising if simpler models tend to converge faster than more complex models.  Nonetheless, it is an important factor to consider when analysing the convergence of SAM predictions, especially in the early universe, where commonly adopted $N$-body simulations (e.g. \citealt{Springel05,Boylan-Kolchin09,Klypin11,Ishiyama21}) tend to have very poor time cadence compared to our SURFS suite (and even compared with Hi-SURFS).

Despite the small differences between the outputs of \shark\ used here resulting from the different $N$-body simulations adopted, our results show that \shark\ out of the box is capable of predicting UVLFs that are in reasonable agreement with observations even up to $z=17$. This marks a stark contrast with other literature results, which show other models struggling with this and needing to invoke modifications to the baryon physics, including the star formation efficiency (e.g. \citealt{Somerville25}) or variations to the IMF (e.g. \citealt{Hutter25,Fontanot26,Durrant26}). 

\subsection{Demonstrating the impact of disk instabilities and more efficient SFE starbursts on the UV luminosity function}

%The dotted lines in Fig.~\ref{UVLF} show the total UV LF of runs s obtained when disk instabilities are not included.  
The results above raise the question of why does \shark\ perform well in this regime, while other models appear to struggle. To understand what the key baryon physics aspects are that allow \shark\ to succeed in reproducing the UVLF in the very early universe, we turn to exploring variations of the fiducial model. 
Fig.~\ref{UVLFnodiskins} shows the predicted UVLF from $z=6$ to $z=17$ in the fiducial \shark\ run using the L800 box (solid lines) and two variations: one 
%Dashedlines in Fig.~\ref{UVLFnodiskins} show the predicted UV LF of our fiducial run and a 
run has identical parameters but turns off disk instabilities (dashed lines); and a second one that assumes no difference in the conversion efficiency of molecular gas to star formation rate in star formation associated with disks and starbursts ($\eta_{\rm burst}=1$; dotted lines). When turning disk instabilities off, we simply avoid doing anything to galaxy disks (both their gas and stellar components) when they are globally unstable. This choice in  practice removes that channel of starbursts in \shark. 
%
%We show this only for \shark\ run over the L800 box, but noting that similar results are found for the other runs. 

Without disk instabilities, the bright-end of the UV LF is systematically underestimated at the highest redshifts, $z\ge 13$. The differences between the fiducial run and a run without disk instabilities, become increasingly larger with increasing redshift, but they do converge by $z=6$.  
%and this underestimation becomes increasingly larger from $z=6$, in which the impact is limited, to $z=17$.
These results demonstrate that in \shark\ v2.0, disk instabilities are vital to reproduce the bright end of the UVLF at the highest redshifts at which there are constraints from the {\it JWST}.
%$6\le z \le 17$, 
Without this mechanism, \shark\ predicts number densities that are up to $\approx 1$~dex lower than our fiducial run, especially at $M_{\rm UV}\lesssim -17$. The results of the run without disk instabilities resemble the results recently reported for other semi-analytic models, such as the Santa Cruz SAM \citep{Somerville25} and GAEA \citep{Fontanot26}. Neither of these models includes a starburst channel associated with globally unstable disks, which may be at the root cause of the difference with our predictions.
%do not trigger central starbursts. 

Assuming a universal conversion efficiency between $\Sigma_{\rm mol}$ and $\Sigma_{\rm SFR}$ also leads to an increasingly dimmer bright-end of the UVLF with increasing redshift. This run does not converge to our fiducial run by $z=6$ and instead it continues to predict a bright-end of the UVLF that is too faint compared with the fiducial run. 
The results of the run with $\eta_{\rm burst}=1$ demonstrate that having additional channels of starbursts is not sufficient to guarantee the prediction of enough bright UV galaxies in the early universe--these starbursts need to be associated with more efficient conversion of molecular gas into star formation. However, such increased efficiency does not need to be more than what observations indicate at lower redshifts, with \shark's $\nu_{\rm boost}$ parameter being informed by $z<2$ observations (e.g. \citealt{Genzel10}). This means that the increased SF efficiency does not need to be outside the range already probed by observations. 

\subsection{On the physical origin of the UV luminosity of high-z galaxies}

Fig.~\ref{UVLFchannels} shows the breakdown of the origin of the UV luminosity as a function of the UV magnitude for the \shark\ v2.0 run with the L800 simulation. The figure shows that the UV brightest galaxies tend to be primarily associated with starbursts originating from disk instabilities. At $z\ge 10$, in fact, the majority of galaxies with $M_{\rm UV}\lesssim -17$ have most of their UV luminosity coming from starbursts driven by disk instabilities, while at fainter magnitudes, star formation in disks becomes the primary mechanism of UV luminosity. In the runs without disk instabilities we find that many of the galaxies that would have experienced instabilities and thus starbursts, have high SFRs associated with star formation in disks, with the resulting UV luminosity only slightly lower than what we get in our fiducial run (not shown here). This reflects the fact that 
%efficient self-regulation of star formation at these high redshifts, with 
star formation in disks would proceed from the gas that would have otherwise been used to trigger a central starburst when disk instabilities are off, indicating an efficient self-regulation of star formation in the early universe. 

Fig.~\ref{UVLFchannels} also shows the contribution from star formation associated to galaxy disks in two of the smaller volume, higher resolution runs. Typically we find that these runs also show a sharp transition from galaxies whose UV luminosity is primarily associated with star formation in galaxy disks, to brighter galaxies dominated by starbursts driven primarily by disk instabilities. This transition moves from brighter UV magnitudes, $\approx -18$ at $z\approx 8$, to fainter ones, $\approx -16.5$, at $z\approx 15$. There is a difference in the predictions of these high resolution boxes and the lower resolution L800 box, where the latter tends to prefer a fainter UV magnitude marking the transition between galaxies dominated by star formation in disks to centrally concentrated starbursts. We note that Hi-SURFS has a much higher resolution than micro-SURFS, but their predictions agree well with each other, indicating that the transition UV magnitude is well converged at the resolution of micro-SURFS. 

The difference between the dominant star formation channel in the fainter vs. brighter UV galaxies in \shark\ should have a clear consequence on the luminosity-weighted sizes of galaxies. In the future, measuring the UV sizes of galaxies in observations across the magnitude range at $z>6$ will help to isolate broadly different star formation driving mechanisms and test our predictions. 

\subsection{On the stochasticity of the UV--stellar and UV--halo mass relations}

\begin{figure*}
\begin{center}
\includegraphics[trim=4.5mm 4.5mm 2mm 2mm, clip,width=0.49\textwidth]{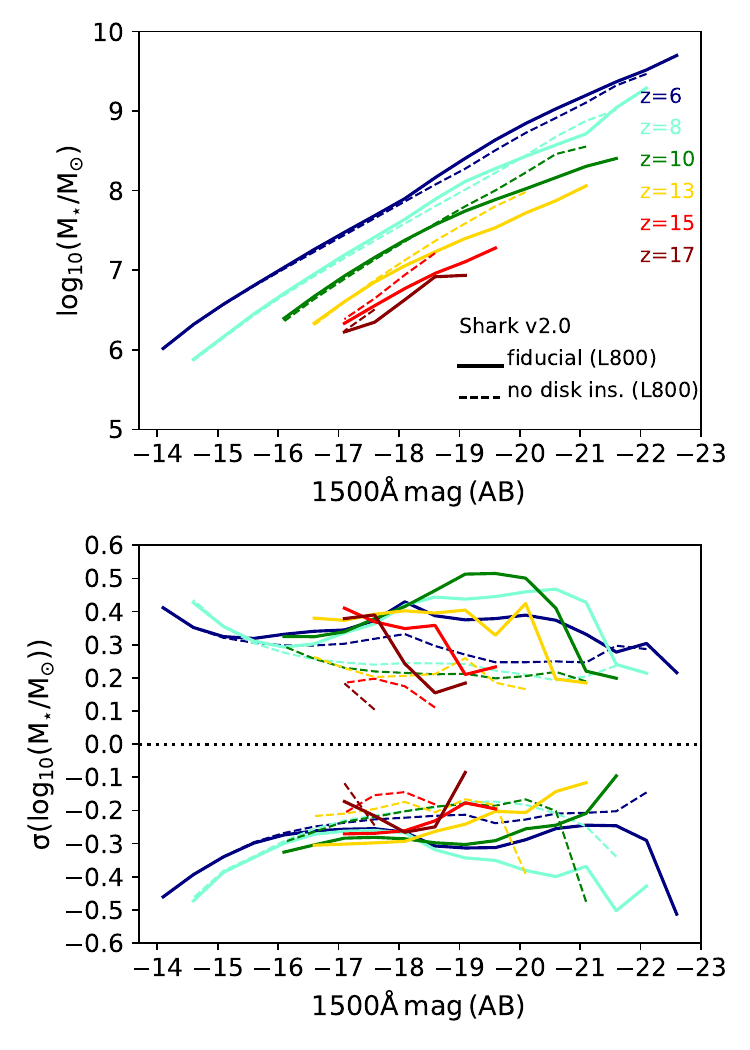}
\includegraphics[trim=4.5mm 4.5mm 2mm 2mm, clip,width=0.49\textwidth]{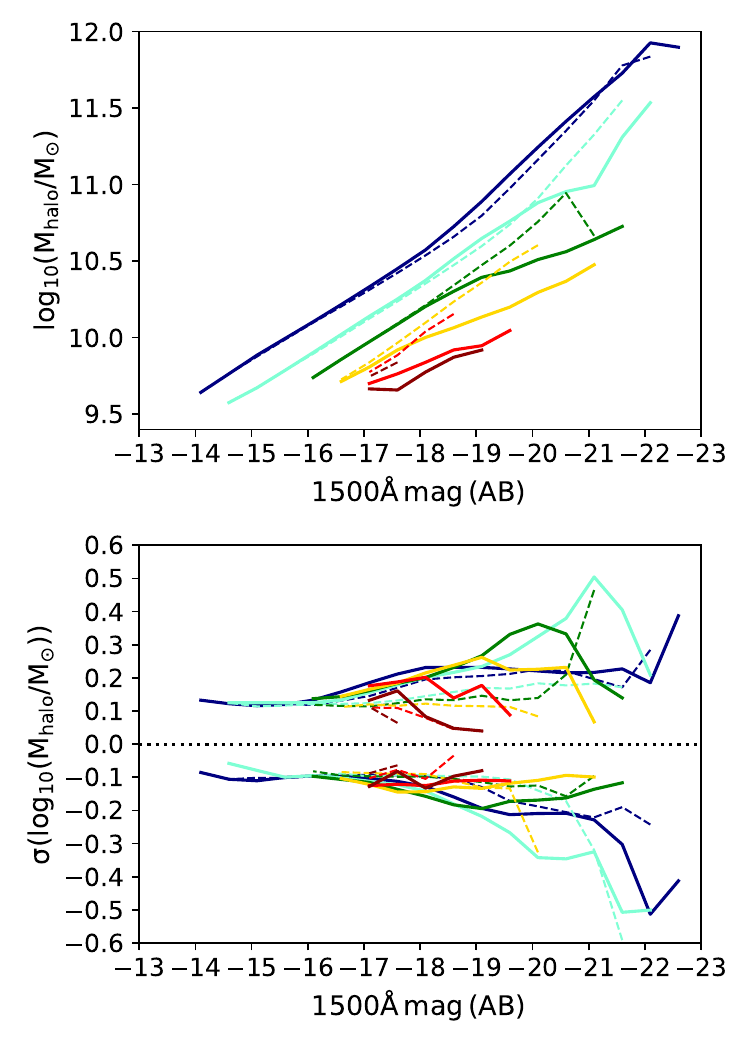}
\caption{{\it Top panels:} The median stellar (left) and host halo (right) mass in bins of the UV magnitude of galaxies in \shark\ from $z=6$ to $z=17$, as labelled in the top-left panel. We show two runs of \shark\ v2.0 on the L800 box: the fiducial run (solid lines) and a run without disk instabilities (dotted lines), as labelled. Only bins with $\ge 10$ galaxies are shown. We only plot the UV magnitude range at which the UV LF in the L800 box is converged (based on comparisons with the two higher resolution runs), which happens at $<-14,\,-14.5,\, -16,\, -16.5,\, -17,\, -17$ at $z=6,\, 8,\, 10,\, 13,\, 15,\, 17$, respectively. {\it Bottom panels:} the scatter in the relations of the top panel, calculated as the distance between the $16^{\rm th}$ and $84^{\rm th}$ percentile ranges of the distribution of galaxies.}
\label{UVscaling}
\end{center}
\end{figure*}

%\begin{figure}
%\begin{center}
%\includegraphics[trim=4.5mm 4.5mm 2mm 2mm, clip,width=0.49\textwidth]{Figs/StellarMass_SFR_zGT6.pdf}
%\caption{{\it Top panels:} The median SFR (top) and $16^{\rm %th}-84^{\rm th}$ percentile ranges (bottom) as a function of stellar %mass for galaxies in \shark\ from $z=6$ to $z=17$, as labelled. The %scatter of the main sequence tends to increase from $z=6$ to $z=13$ %at fixed stellar mass, especially related to the up-scatter (above %the main sequence), while also increasing with increasing stellar %mass at fixed redshift. }
%\label{mainseqscaling}
%\end{center}
%\end{figure}

Beyond demonstrating that \shark\ without disk instabilities would lead to a bright end of the UVLF that is discrepant with observations at $z>10$, we want to understand how the inclusion of disk instabilities impacts the relationship between UV magnitude and stellar and halo mass. 
%the direct impact of disk instabilities 
The left panels of Fig.~\ref{UVscaling} show the median and scatter around the relationship between the rest-frame UV absolute magnitude and stellar mass. For clarity, we only show this for \shark\ using the L800 box. We show this for our fiducial run (solid lines) and a run without disk instabilities (dotted lines). 

There are several important trends worth highlighting: (i) the median stellar mass associated with a fixed UV magnitude monotonically decreases with increasing redshift; (ii) the scatter in the UV magnitude-stellar mass relation tends to increase at both ends of the UV magnitude range, with a minimum that sits between $\approx -17$ and $\approx -18$, depending on the redshift; 
%peak at $M_{\star}\approx 10^{7}-10^{8.5}\,\rm M_{\odot}$; 
and (iii) the scatter tends to increase with increasing redshift, which is especially clear around $M_{\rm UV}\approx -18$ to $-20$ and between $z=6$ and $z=13$. At higher redshifts, statistics become quite limited, and thus it is hard to assert whether there is a reversal or not of the evolution of the scatter. We bring the attention of the reader to the fact that the slope of the relation between stellar mass and UV magnitude is not evolving in a significant manner, and instead the redshift evolution appears to be a monotonic decrease of the zero-point of the relation with increasing redshift. Thus, at fixed stellar mass, galaxies can be much brighter in UV at $z>13$ than they are at $z\approx 6$. The fixed slope reflects the underlying main sequence evolving only in normalisation, suggesting self-regulation continues to be effective at these very high redshifts in \shark. 

When the fiducial \shark\ run is compared with the run without disk instabilities, we find that at $z\ge 8$, the brightest UV galaxies are associated with higher mass galaxies in the run without disk instabilities than in the fiducial run; in other words, galaxies need to grow further in stellar mass to be as bright in UV in the absence of a starburst channel associated with disk instabilities. This naturally explains why that model predicts a fainter end of the UVLF at those high redshifts. In tandem with this effect, we find that there is significantly larger variance in the stellar mass range associated with a fixed UV magnitude in the fiducial model compared with the run without disk instabilities, indicating significantly higher variance when disk instabilities are on. This is particularly strong when we focus on the up-scatter (the $84^{\rm th}$ percentiles), with differences in the down-scatter  (the $16^{\rm th}$ percentiles) between the two models being more modest. This increased ``stochasticity'' has been invoked ad-hoc in empirical models to reproduce the UVLF at $z>6$ \citep{Gelli24}. \shark\ demonstrates that this can naturally arise from a combination of frequent starbursts associated with disk instabilities, and an overall higher $\Sigma_{\rm mol}-\Sigma_{\rm SFR}$ conversion efficiency during starbursts. 
%for the runs at $z\ge 8$, while there is a clear increase in the scatter at all stellar masses at $z=6$ relative to the higher redshift galaxies.

The right panels of Fig.~\ref{UVscaling} show the median and scatter around the relationship between UV magnitude and halo mass. This time we limit the analysis to central galaxies only. Although satellites rarely contribute, we find that at the lowest redshift analysed ($z=6$), they become dominant enough at the high-halo mass end ($M_{\rm halo}\gtrsim 10^{11.5}\,\rm M_{\odot}$) as to skew the median and scatter. 

The relationship between UV magnitude and halo mass is very steep, with the median $M_{\rm UV}$ spanning $\approx 5$ magnitudes in the halo mass range $10^{9.5}-10^{10.5}\,\rm M_{\odot}$ at $z\approx 6-8$. We note that the slope of the relation between halo mass and $M_{\rm UV}$ becomes shallower with increasing redshift, with a similar range of UV magnitude spanning an increasingly smaller range of halo mass as we move from $z=6$ to $z=17$.

We also find the same increasing median UV magnitude with increasing redshift at fixed halo mass that we described for stellar mass, but this time the slope of the relation is evolving noticeably. For instance, at $z=13$, the range of UV magnitudes $[-16.5,-21]$ is associated with halo masses in the range $\approx 10^{9.7}-10^{10.5}\,\rm M_{\odot}$, while the same range of 
UV magnitudes at $z=6$ spanning a halo mass range of $10^{10.2}-10^{11.5}\,\rm M_{\odot}$. Interestingly, the scatter between the UV magnitude and halo mass is smaller than that with stellar mass, and it monotonically increases with increasing UV luminosity (while for stellar mass it tends to flare at both magnitude ends). This shows that at these very high redshifts, the UV magnitude appears to correlate more strongly with halo mass than stellar mass (as indicated by the reduced scatter of the correlation). We note that the UV magnitude regime showing the flare in scatter in the halo mass-UV magnitude relation is the regime that is dominated by starbursts being driven by disk instabilities. 
In contrast, at the faint end, the model predicts a 
%while the fainter end has a 
non-evolving magnitude of the scatter. This is the regime that  
%
%dependence of the scatter on halo mass varies with redshift. 
%At $6\le z\le 10$, the scatter increases with increasing halo mass, from values $\approx 0.7-1$~mag at $M_{\rm halo}\approx 10^{9.5}\,\rm M_{\odot}$ to $\gtrsim 2$~mag at $M_{\rm halo} \gtrsim 10^{11.5}\rm M_{\odot}$. However, at $z\ge 13$, the scatter, especially the scatter below the median, tends to increase with decreasing halo mass, with also a clear redshift evolution, with the scatter being larger at $z=17$ than at $z=13$ by $\approx 0.5$~mag. This is the regime that 
is firmly dominated by star formation in disks rather than starbursts in \shark, with the scatter likely driven by stellar feedback. 

%We show in the right panels of Fig.~\ref{UVscaling} the relation  between $M_{\rm UV}$ and halo mass. 
%At halo masses $\lesssim 10^{11}\,\rm M_{\odot}$, 

Once again, comparing the fiducial \shark\ run with the run without disk instabilities, we see two important differences worth highlighting. The first one is that, at $z\ge 8$, the brighter UV galaxies are hosted in lower-mass halos in the fiducial run compared with the run without disk instabilities, similar to what was found for stellar mass. The second one is the reduced scatter around the $M_{\rm UV}$-halo mass relation in the absence of disk instabilities. In fact, we see that the runs without disk instabilities are characterised by an almost constant scatter along the $M_{\rm UV}$ range at $z\ge 8$, which contrasts the flare in the scatter that is obtained in the fiducial run at the bright-end. This shows that disk instabilities are responsible for the increase in variance, which is at the core of why our fiducial \shark\ model reproduces the UVLF evolution at $z>6$. 
%shows a distinct increase towards higher halo masses. At any one redshift, the scatter increases
%oth towards lower and higher halo masses, with a mini at fixed redshift, and towards higher redshifts at fixed halo mass. The scatter reaches even $\gtrsim 1$~mag at $M_{\rm halo}\lesssim 10^{10}\,\rm M_{\odot}$ and $z\gtrsim 13$. 

The above results show that the success of the fiducial \shark\ model in reproducing the UV LF at $z>6$ is a combination of the increasing median UV luminosity associated with a given halo mass (relative to models without disk instabilities) and the enhanced variance of the relation, particularly at the high halo-mass end. In other words, \shark\ reproduces the observed UVLF, especially at the bright-end, in the early universe, because the model is characterised by an increase in stochasticity of the UV luminosity-halo mass relation at the high luminosity-high halo-mass end.  
%
%level of UV luminosity stochasticity in \shark\ becomes more significant at low halo masses and in the very early universe. 
This increased stochasticity at high halo masses contrasts with the net effect obtained in the FIRE-II simulations \citep{Sun23}, where the cumulative effect of their subgrid physics manifests in a SFR stochasticity that increases with decreasing halo mass at $z>10$. It is this increased stochasticity at the low halo-mass regime that allows FIRE-II to reproduce the observed range of UV magnitudes in the very early Universe. Similarly, \citet{Gelli24} invoked a variance in the $M_{\rm UV}$ of galaxies that increases with decreasing host halo mass to demonstrate that such behaviour can help reproduce the UV LF at $z>10$. In contrast, we show here that \shark\ achieves this agreement with the observed UVLF by increasing the stochasticity in the high-mass end. 

The conflicting results above thus indicate that there are likely numerous ways in which a galaxy formation model can reproduce the high-z UVLF, and that additional observations are needed to disentangle competing solutions. One key missing piece of the puzzle right now is environmental constraints of bright UV galaxies. Measurements, such as galaxy clustering, will help disentangle these competing solutions. In addition, models need to be contrasted with observations from the lower redshift universe, which tend to be more robust against systematic effects and also help to probe different regimes of galaxy evolution.

\section{The stellar mass function and cosmic star formation rate density evolution at $0\le z\le 15$}\label{allcosmicepoch}

\begin{figure*}
\begin{center}
\includegraphics[trim=2mm 2mm 2mm 2mm, clip,width=0.99\textwidth]{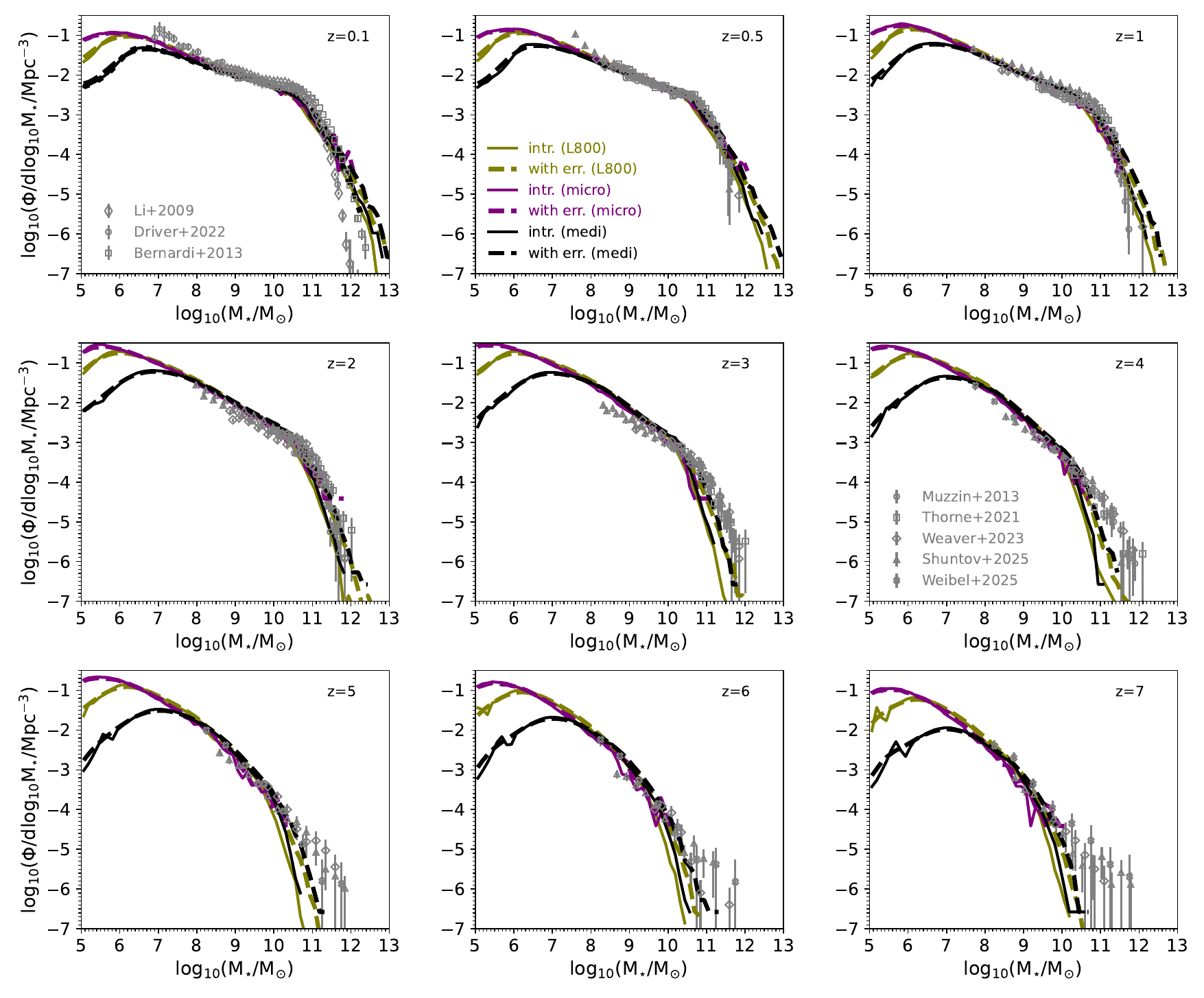}
\includegraphics[trim=2mm 2mm 2mm 2mm, clip,width=0.99\textwidth]{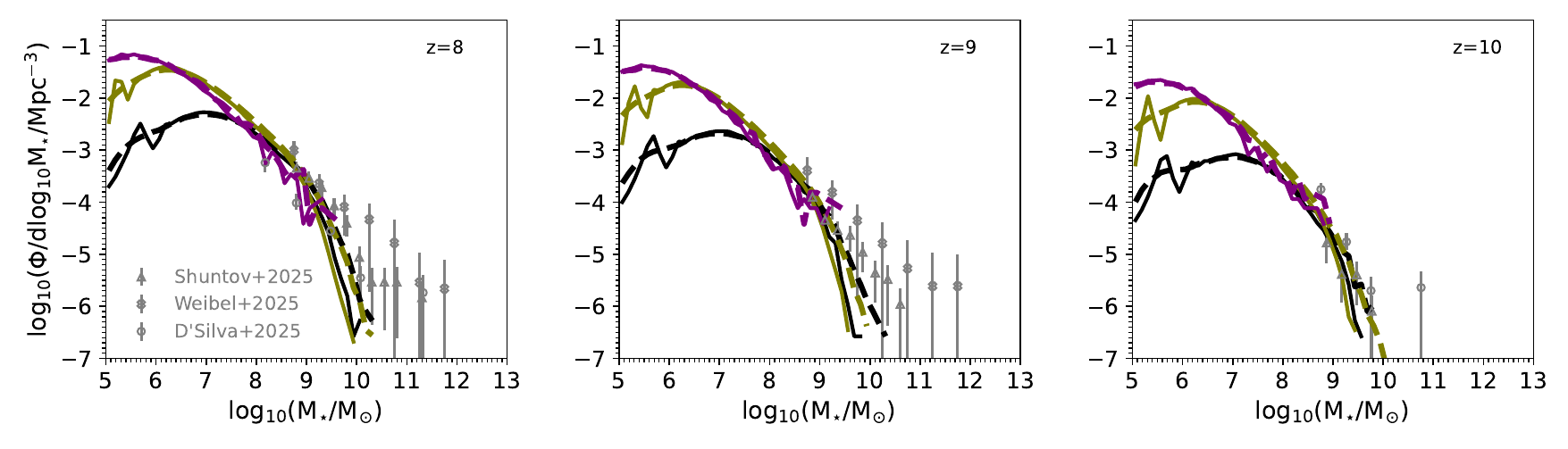}
\caption{Stellar mass function from $z=0.1$ to $z=10$ in \shark\ v2.0 using the medi-SURFS, micro-SURFS and L800, as labelled. For each run we show the intrinsic measurements (using the true stellar masses; solid lines) and the measurements after we convolve with a Gaussian-distributed error (dashed lines; Eq.~\ref{errsm}). Observational measurements from \citet{Li09,Driver22,Bernardi13,Muzzin13,Thorne21,Weaver22,Shuntov25,Weibel24b,DSilva25}, are shown as symbols, as labelled. } 
\label{SMF}
\end{center}
\end{figure*}

\begin{figure*}
\begin{center}
\includegraphics[trim=2mm 4mm 2mm 3mm, clip,width=0.99\textwidth]{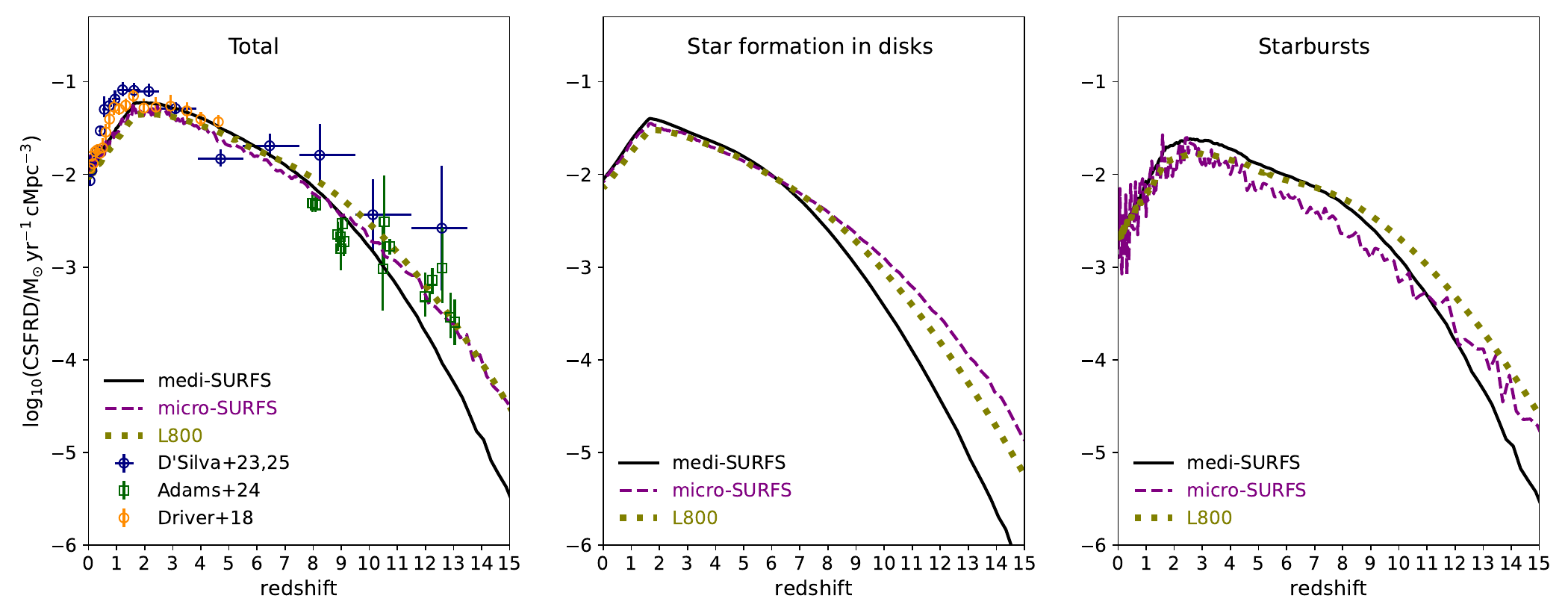}
\includegraphics[trim=2mm 4mm 2mm 3mm, clip,width=0.99\textwidth]{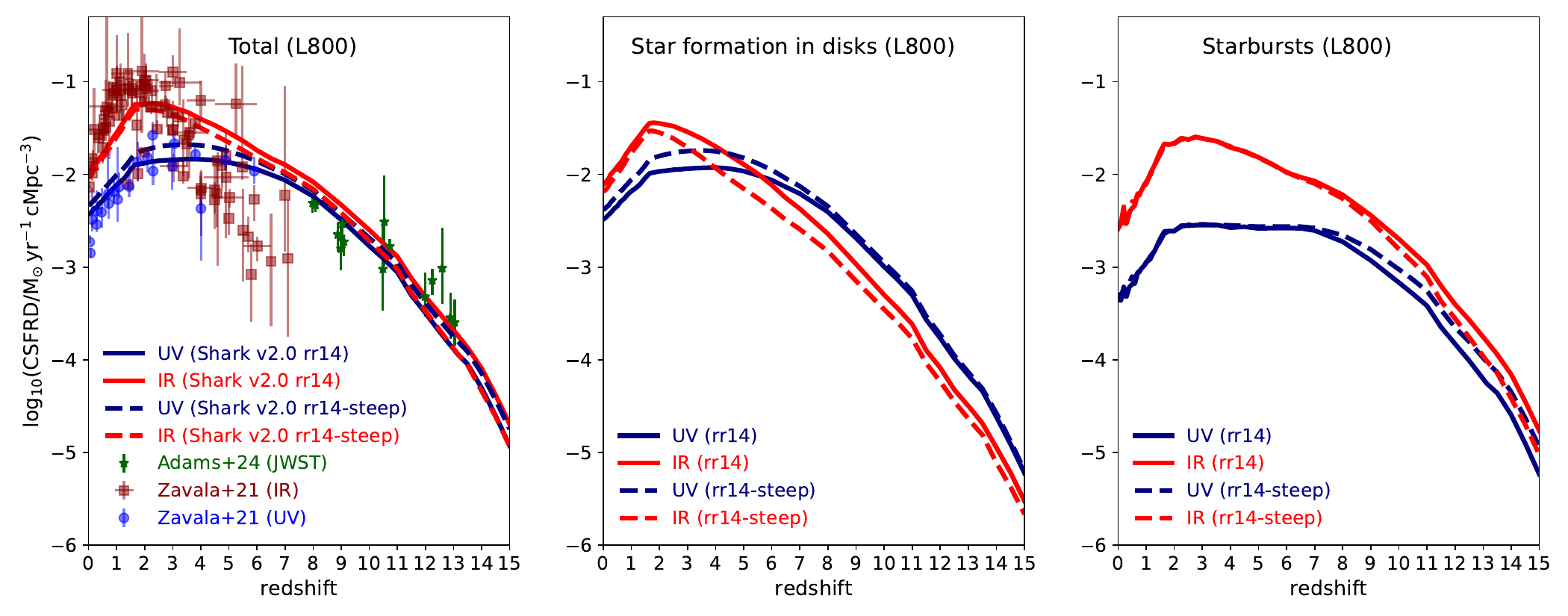}
\caption{{\it Top panels:} Evolution of the CSFRD at $0\le z\le 15$ in \shark\ v2.0 using medi-SURFS, micro-SURFS and L800, as labelled. This is shown for the total SFR of all galaxies (left panel), and for the contributions of star formation in disks (middle panel) and starbursts (right panel). We show observational constraints on the left panel from \citet{Driver17,D'Silva23,DSilva25,Adams23}, as labelled. The data shown as ``Adams+24'' corresponds to a compilation of data plus that presented in that manuscript, including data from \citet{Perez-Gonzalez23,Donnan23,Harikane23,Finkelstein24,Willott24,McLeod24}. Observational data at $z>6$ correspond to {\it JWST} results. {\it Bottom panels:} As in the top panels but for \shark\ v2.0 run over L800 only. We show separately the obscured (red lines) and unobscured (blue lines) contributions to the CFRD. The solid and dashed lines show two different metals-to-dust scaling, both from \citet{Remy-Ruyer14} (see \S~\ref{dustsec} for a description). For the total CSFRD, \shark\ v2.0 prefers an unobscured contribution similar or only slightly above the obscured CSFRD at $z\gtrsim 9$. We show an observational compilation of the obscured (square symbols) and unobscured (circles) CSFRD, and the \citet{Adams23} compilation of JWST results also shown in the top panel. The obscured CSFRD data come from \citet{Madau14,Cucciati12,Casey12,Casey21,Zavala21}, while the unobscured CSFRD data come from \citet{Cucciati12,Reddy09,Bouwens12,Robotham11,Finkelstein15,Dahlen07,Schenker13}.} 
\label{cosmicsfr}
\end{center}
\end{figure*}

One of the key challenges that galaxy formation models and simulations face is to simultaneously reproduce a variety of observations that span a wide range of cosmic epochs. In this section, we demonstrate that the model used in \S~\ref{uvlfsec} to compare \shark\ predictions with observations of the UVLF at $z>6$, also provides a reasonably good fit to the stellar mass function and CSFRD evolution down to $z=0$.

\subsection{The stellar mass function}
Fig.~\ref{SMF} shows the stellar mass function from $z=0.1$ to $z=10$ in \shark\ using the simulation boxes listed in Table~\ref{tab:sims} that reach $z=0$. We note that the $z=0.1$ SMF was used to fit the fiducial model--it is one of the inputs of the PSO {\sc optim} package of \shark--as described in \citet{Lagos24}. However, higher redshifts were not used as inputs to the PSO. 

We show two sets of predictions, the SMF calculated from the intrinsic stellar masses obtained in \shark, and what the model predicts after convolving the intrinsic stellar masses in \shark\ with a Gaussian-distributed error with a mean of $0$ (i.e. no bias) and a standard deviation, $\sigma_{\rm SM}$ in $\rm log_{10}(M_{\star})$, that is redshift-dependent. For the latter, we follow \citet{Behroozi19,Chaikin26}, defining:

\begin{equation}
    \sigma_{\rm SM}= {\rm min}\left[ \sigma_{\rm r,0} + \sigma_{\rm r,z}\,z, \sigma_{\rm r,max}\right],\label{errsm}
 \end{equation}

\noindent where $\sigma_{\rm r,z}=\sigma_{\rm r,0} = 0.1$, $\sigma_{\rm r,max}=0.3$, and $z$ is the redshift. 

Fig.~\ref{SMF} shows that by combining the three different simulations in Table~\ref{tab:sims}, we can predict the SMF in a very wide dynamic range, from $M_{\star}\approx 10^{5.7}-10^{12.5}\rm \, M_{\odot}$. In the regime in which each simulation is well converged, we see excellent agreement between \shark\ runs, regardless of the different underlying methods to build halo catalogues and merger trees, particularly between L800 and the \surfs-simulation suite. We remind the reader that we use exactly the same models and parameters for each of these runs. 
The excellent convergence is the result of the methods introduced in \citet{Chandro-Gomez25a} successfully mitigating numerical artefacts arising from the halo finding and merger tree building, which particularly affect the L800 halo and subhalo catalogue.

Regarding the convergence between simulations, it is interesting to note that a small difference in resolution, for example between the L800 and medi-SURFS (i.e. a factor of $\approx 2.5$ in the minimum halo mass; see Table~\ref{tab:sims}) can yield differences in the sampled low-mass end of up to $\approx 1.5$ orders of magnitude. This is due to the steepness of the stellar-halo mass relation at the low-mass end, which is well documented in the literature (e.g. \citealt{Behroozi13,Moster13}) and that \shark\ reproduces well (see \citealt{Lagos18c,Lagos24}).

Another important finding in Fig.~\ref{SMF} is that \shark\ provides an excellent overall match to the large set of observations presented in Fig.~\ref{SMF}. The main caveat of that statement is that the model may be under-predicting the abundance of very massive galaxies, $M_{\star}\gtrsim 10^{11}\,\rm M_{\odot}$, between $3\lesssim z \lesssim 7$. Most of these galaxies are star-forming, dusty galaxies in \shark. \citet{Mitchell13} showed that deriving stellar masses for those galaxies in observations from SED fitting can be particularly challenging, with the derived SMF being biased high at the very high-mass end relative to intrinsic values at $2\lesssim z\lesssim 4$ (see their fig.~12). \citet{Long22} in fact argue that $50-100$\% of galaxies with stellar masses $\gtrsim 10^{11}\,\rm M_{\odot}$ are dusty, star-forming galaxies at $z>1$. Thus, we consider the current level of tension acceptable given the uncertainties on the observational side. 

Note that at $z\gtrsim 7$, there are significant differences between observational estimates that can be $\gtrsim 1$~dex in normalisation. This stems from the uncertainty around Little Red Dots (LRDs) and their stellar mass estimates. One can see \citet{Weibel24b} and \citet{Shuntov25} as bracketing the extreme cases in which LRDs are treated in the same way as other galaxies \citep{Weibel24b} or where they are removed from the galaxy sample \citep{DSilva25,Shuntov25}. In general, \shark's predictions are closer to the estimates of \citet{DSilva25} and \citet{Shuntov25}.

\subsection{The cosmic star formation rate density}

For the CSFRD (Fig.~\ref{cosmicsfr}), we find that \shark\ is in very good agreement with current observational constraints at $z\lesssim 6$. There is a tendency for the \shark's CSFRD to peak at slightly higher redshift, $\approx 2$, than observations prefer ($z\approx 1.5-2$).      

At $z>6$, there is considerable uncertainty in the observations, with the predictions of \shark\ lying between the current uncertainties. It is remarkable though that it is only with high-enough resolution (micro-SURFS and L800) that the models predicts a normalisation of the CSFRD as high as the observations indicate.
%the higher normalisation that follows closer the higher observational values is achieved. 
This is a reflection of the significant contribution to the CSFRD at $z\gtrsim 11$ from low-mass halos $\lesssim 4 \times 10^{9}\, {\rm M}_{\odot}\,h^{-1}$. The middle and right panels of Fig.~\ref{cosmicsfr} show the breakdown of SFR channels in \shark. It is interesting to see that the starburst contribution to the CSFRD does not converge for the small volume runs (i.e. micro-SURFS) with a smaller contribution from starbursts compared to the L800 run. The cosmological volume of L800 is $\approx 2,500$ times larger than micro-SURFS, which helps to capture the rare events that can potentially have a significant contribution to the CSFRD, thus minimising cosmic variance. We see that the higher resolution of micro-SURFS and L800 relative to medi-SURFS helps to increase both the SFR associated with disks and starbursts, indicating that both of these star formation channels are prevalent in low-mass halos at these very high redshifts in \shark. 

In the bottom panels of Fig.~\ref{cosmicsfr} we show the CSFRD derived from the IR and UV luminosity of galaxies in \shark. This is more directly comparable to what is measured in observations. To calculate the obscured (unobscured) contributions to the CSFRD in \shark, we sum the total IR (UV) luminosity of all galaxies at a fixed redshift and convert that to a SFR using the \citet{Kennicutt12} conversion factors, which are the ones commonly adopted in the literature. This allows us to carry out a much fairer comparison with observations than we could do by directly taking the instantaneous SFRs of galaxies predicted by \shark. 

We follow the process above for the two dust-to-metal scaling models introduced in \S~\ref{dustsec}. We then compare with the compilation of observations presented in \citet{Zavala21}, and listed individually in the caption of Fig.~\ref{cosmicsfr}. Overall, we see remarkable agreement between \shark\ and the observations of the obscured and unobscured CSFRD at $z\lesssim 4$. At higher redshifts, the contribution from the obscured CSFRD to the total CSFRD is highly debated and potentially impacted by the scarcity of large area millimetre surveys (see \citealt{Casey21} for a discussion). What is clear, however, is that \shark\ predicts that the contribution from obscured star formation continues to be important even up to $z=15$. Depending on the exact dust-to-metal scaling we adopt, we find that obscured star formation can contribute $\approx 25-50$\% of the total CSFRD. 

The middle and left bottom panels in Fig.~\ref{cosmicsfr} show the breakdown of the IR and UV CSFRD for star formation associated with disks and starbursts, respectively. We find that galaxy disks have most of their star formation unobscured at $z\gtrsim 3$, with the exact transition redshift depending on the adopted dust-to-metal scaling. Conversely, for starbursts, it is only at $z\gtrsim 13$ that most of the star formation is unobscured, and only in one of the adopted dust-to-metal scalings (rr14-steep). This stems from the fact that the surface density of gas and metals in starbursts is much higher than that in galaxy disks, which in the model described in \S~\ref{dustsec} directly translates into a higher surface density of dust. Again, measurements of galaxy sizes in different wavelengths at $z>6$ would greatly help to verify our predictions.

Overall, the results shown here build a picture in which it is possible to simultaneously reproduce the very early Universe's observations while also reproducing the more robust observations of the later Universe, with a consistent star formation model, which is inspired by local Universe observations of resolved molecular and atomic gas reservoirs in galaxies, and a stellar feedback model that is informed by analytic models of the evolution of SNe-inflated bubbles in a multi-phase ISM. 
%say something about the difficulty in measuring stellar masses under the significant presence of dust \citep{Mitchell13.}

%\section{Discussion \& Conclusions}\label{conclusions}

%Discuss how realistic the treatment of disk instabilities is. It's certainly a toy model, but what is clear is that it's leading to the boost in stochasticity and starbursts required to reproduce the observations. 

%Also highlight the boost in SFE and its importance. 

%Discuss the impact of dust attenuation and the simple approach we have taken.

%Discuss the issue of resolution (in terms of halo mass range probed but also in terms of time cadence of the outputs). 

%Discuss how we obtain a UVLF in agreement with observations compared to what other papers have focused on. 

%Stress the fact that the model provides a single view of the universe' evolution, with consistent physics across all redshifts.

\section{Discussion}\label{discussion}

Our results demonstrate that the abundance of UV-bright galaxies observed by {\it JWST} at $z>10$ can be reproduced within the same galaxy formation framework that describes galaxies at much later cosmic times, without recalibrating the model or introducing additional redshift-dependent baryonic physics (Fig.~\ref{UVLF}). The physical origin of this agreement in \shark\ is not simply an overall enhancement of the amount of star formation at early times. Rather, it arises from the combination of starbursts triggered by globally unstable disks, the enhanced star formation efficiency adopted during these bursts, and the resulting increase in the stochasticity of UV luminosity at fixed stellar and halo mass. Here we discuss the physical interpretation of these results, the limitations of the modelling, and how the mechanisms operating in \shark\ compare with other proposed explanations for the abundance of UV-bright galaxies during cosmic dawn.

\subsection{Violent disk instabilities, starbursts and stochasticity}

A central result of this work is that the treatment of globally unstable disks in \shark\ plays a critical role in reproducing the observed bright end of the UVLF at $z\gtrsim10$ (Fig.~\ref{UVLFnodiskins}). In the fiducial model, a globally unstable disk transfers its gas and stars into a spheroidal component, with the inflowing gas triggering a starburst. At $z\gtrsim10$, these instability-driven starbursts dominate the UV luminosity of most galaxies brighter than $M_{\rm UV}\approx-17$, while fainter galaxies are predominantly powered by star formation in disks. Removing this channel reduces the abundance of bright galaxies by up to $\approx 1$~dex, with the difference relative to the fiducial model increasing towards higher redshift.

%It is important, however, not to interpret this result as evidence that the specific disk-instability prescription adopted in \shark\ provides a detailed description of the dynamics of galaxies at $z>10$. The instability criterion used here is necessarily highly idealised. Galaxies are represented by disk and spheroid components and a global stability criterion determines when the entire disk becomes unstable, after which its material is transferred to the spheroid. The complex gas dynamics, fragmentation, clump migration and radial inflows expected in extremely gas-rich high-redshift galaxies cannot be followed explicitly by a semi-analytic model. In this sense, the violent disk-instability prescription should be regarded as an effective, or ``toy'', representation of a potentially much richer set of processes capable of rapidly funnelling gas into dense star-forming regions.

The treatment of violent disk instabilities in \shark\ is necessarily idealised, and should not be interpreted as a detailed description of the dynamics of galaxies at $z>10$. Nevertheless, the qualitative outcome of the model, namely, rapid gas inflow followed by compact and intense star formation, is increasingly supported by hydrodynamical simulations of galaxies during cosmic dawn. \citet{Shen24} using the THESAN simulations, find that massive galaxies at $z\gtrsim 6$ undergo rapid compaction and gas depletion, which is likely driven by gravitational instabilities within gas disks and is accompanied by centrally concentrated star formation. Similarly, \citet{Roper23} using FLARES, predict a high-redshift regime dominated by compact-core star formation, with some galaxies undergoing runaway central star formation as efficiently cooling gas reaches high densities. Most recently, \citet{Cataldi26} using the FirstLight simulations, have revealed a wet-compaction phase at $z>6$, in which self-reinforcing gas inflows trigger strong, localised starbursts within the central $\approx 1$~kpc. In the latter, galaxies in the massive end are more likely to undergo this compaction mechanism, similar to the mass range at which disk instabilities in \shark\ become dominant (Fig.~\ref{UVLFchannels}). 
Although the physical drivers of these events differ between simulations, these results support a picture in which rapid gas inflows and centrally concentrated starbursts are a natural feature of sufficiently massive galaxies during cosmic dawn.

The phenomenological consequence of violent disk instabilities in \shark\ on the statistics of star formation is robust. Removing disk instabilities does not simply suppress star formation altogether: much of the gas that would have participated in an instability-driven burst subsequently forms stars in the disk. The decisive difference is, instead, that the fiducial model concentrates this star formation into shorter and more efficient episodes. Consequently, disk instabilities both increase the median UV luminosity at fixed halo mass and substantially broaden the $M_{\rm UV}$-halo mass relation at its bright end. Without disk instabilities, this relation has an approximately constant scatter at $z\gtrsim8$; with them, the scatter increases strongly towards brighter galaxies. It is this enhanced stochasticity that populates the bright tail of the UVLF.

This interpretation connects our results to a broader body of work emphasising burstiness as a possible explanation for the high abundance of UV-bright galaxies (e.g. \citealt{Sun23,Gelli24,Semenov25}). There are, however, important differences in where this stochasticity arises. FIRE-II predicts increasing SFR stochasticity towards decreasing halo mass \citep{Sun23}, while \citet{Gelli24} similarly explored a scatter in UV luminosity that increases towards lower halo masses. In \shark, the behaviour required to reproduce the UVLF instead emerges most strongly at the high-luminosity, high-halo-mass end as a consequence of instability-driven starbursts. Thus our results show that agreement with the UVLF alone does not uniquely identify the underlying mechanism. Measurements sensitive to the host halo masses and spatial distribution of UV-bright galaxies, particularly clustering, will be important for distinguishing these scenarios. The predicted transition from disk-dominated UV emission at faint magnitudes to compact, starburst-dominated emission at brighter magnitudes also suggests that resolved UV sizes and morphologies provide an independent observational test.

\subsection{Enhanced star formation efficiency during starbursts}

Triggering starbursts alone is not sufficient. When we set $\eta_{\rm burst}=1$, such that molecular gas forms stars with the same efficiency in disks and starbursts, the bright end of the UVLF becomes progressively fainter relative to the fiducial model towards high redshift (Fig.~\ref{UVLFnodiskins}). Unlike the model without disk instabilities, this difference persists even at $z=6$. The success of \shark\ therefore relies on both the triggering of burst events and the faster conversion of molecular gas into stars during those events.

This result has an interesting connection to models in which an enhanced star formation efficiency at high redshift has been introduced explicitly to explain the JWST observations (e.g. \citealt{Somerville25}). In \shark, however, the enhancement is neither introduced specifically at high redshift nor applied universally to the galaxy population. The factor $\eta_{\rm burst}=15$ is applied only to molecular gas participating in starbursts and was adopted before the {\it JWST} observations, motivated by the shorter molecular-gas depletion times measured in starbursting galaxies at substantially lower redshifts (e.g. \citealt{Daddi10,Genzel10,Genzel15,Scoville16,Tacconi18}). The high-redshift evolution instead emerges because conditions in the early Universe cause an increasing fraction of UV-bright galaxies to enter this high-efficiency mode.

This distinction is important. \citet{Somerville25} demonstrated that sufficiently high star formation efficiencies can reproduce the observed $z>10$ UVLF, but extreme efficiencies can overproduce galaxies at later epochs. In \shark, enhanced efficiency is associated with a particular physical state rather than imposed as a general redshift-dependent modification. The same prescription can therefore generate highly efficient episodes of star formation when required at early times while retaining the much lower effective efficiencies needed to reproduce the galaxy population at later epochs. Our results suggest that the relevant question may therefore not be whether star formation is universally more efficient at cosmic dawn, but rather whether the physical conditions that generate short-lived, high-efficiency modes of star formation become sufficiently common.

\subsection{The role and uncertainty of dust attenuation}

Dust attenuation remains an important uncertainty when connecting intrinsic star formation to the observed UVLF. In \shark, dust masses are inferred from the gas mass and metallicity using locally calibrated dust-to-metal relations, while attenuation is calculated from the resulting dust surface density using a two-component birth-cloud and diffuse-ISM model, informed by radiative transfer calculations in cosmological hydrodynamical simulations \citep{Trayford20}. This treatment has the advantage of coupling attenuation to the evolving physical properties of model galaxies rather than prescribing an explicit redshift evolution. The predicted dust-to-stellar mass ratios also agree reasonably well with the observational constraints currently available to $z\sim7$ (Fig.~\ref{mdustmstar}).

Nevertheless, extrapolating these relations to the first few hundred Myr should be treated cautiously. At these epochs, dust production, destruction and grain growth may occur on timescales comparable to the ages of the galaxies themselves, and our model does not explicitly follow these processes. This is particularly relevant given that delayed grain growth has been shown to affect the high-redshift UVLF predicted by GALFORM \citep{Lu25}.  Appendix~\ref{dustatt} demonstrates the impact of the two dust-to-metal scaling described in Eqs.~\ref{eq.mdust}~and~\ref{eq.mdust2} on the attenuated UVLF in \shark, showing that there are systematic differences between the two even at $z=17$. Our two adopted dust-to-metal relations provide some indication of the systematic uncertainty associated with the amount of dust, but they cannot encompass the full uncertainty in high-redshift dust physics. 

The impact is also strongly redshift and luminosity dependent. By $z\sim13-17$, attenuation of the UVLF is small in our predictions, whereas at $z=10$ dust can shift the bright end by up to $\sim1.5$~mag or more depending on the adopted dust-to-metal ratio. Moreover, the starbursts responsible for the brightest UV galaxies have much higher gas and metal surface densities than galaxy disks and can therefore remain significantly obscured to very high redshift. Thus, while uncertainties in dust modelling cannot explain the fundamental difference between the fiducial and no-instability \shark\ models at the highest redshifts, they remain important for quantitative predictions of the bright UV population and of the division between obscured and unobscured cosmic star formation.

\subsection{Numerical resolution and time resolution}

The very rapid assembly of galaxies during cosmic dawn also places unusually stringent requirements on the numerical backbone used by semi-analytic models. Our comparison of several $N$-body simulations illustrates two distinct aspects of this problem.

First, mass resolution determines the faintest galaxies that can be reliably modelled. The steep stellar-to-halo mass relation at low halo masses means that relatively modest differences in halo mass resolution translate into much larger differences in the range of observable galaxy properties that can be sampled. For example, the factor of only $\approx 2.5$ difference in particle mass between medi-SURFS and L800 translates into a difference of more than two magnitudes in the UVLF at $z\sim13$. Reaching $M_{\rm UV}\sim-13$ requires resolving halos approaching $10^8\,{\rm M_\odot}$. Similarly, our CSFRD predictions show that halos with masses below $\sim 4\times10^9\,{\rm M_\odot}\,h^{-1}$ make an important contribution at $z\gtrsim 11$. Predictions for both the faint UVLF and the total star formation budget at these epochs are therefore particularly sensitive to mass resolution.

Second, temporal resolution is itself important. Halos grow extremely rapidly at $z>10$, while the starbursts responsible for the UV-bright population occur over short timescales. Hi-SURFS has substantially better mass resolution than several of the other simulations considered here, yet produces systematically different UVLFs because its snapshot cadence is approximately a factor of two poorer. Our experiments suggest that $\gtrsim 70$ snapshots at $z>6$ are required for convergence of the UVLF in the version of \shark\ employed here. This should not be interpreted as a universal numerical requirement: the necessary cadence will likely depend on the galaxy formation model and, particularly, on how rapidly its baryonic reservoirs respond to changes in halo properties. It does, however, demonstrate that high mass resolution alone is insufficient for modelling cosmic dawn. Semi-analytic predictions at these epochs also require merger trees capable of resolving the very short timescales on which haloes and galaxies evolve. This is a significant and much less explored area of convergence in SAMs. In fact, most of the widely adopted $N$-body simulations, such as Millennium and Millennium-II \citep{Springel05,Boylan-Kolchin09}, and Uchuu \citep{Ishiyama21} have a much poorer time cadence, with $<70$ snapshots between the start of the simulation and $z=0$. Bolshoi \citep{Klypin11} on the other hand has $180$ snapshots to $z=0$, however, only $26$ are to $z=6$. This is much worse than the simulations used in this work (see Table~\ref{tab:sims}), demonstrating that the axis of time cadence is as important as the mass resolution when simulating the high-z universe with SAMs.  

\subsection{Why does \shark\ reproduce the high-redshift UVLF?}

A variety of solutions have now been proposed to explain the abundance of UV-bright galaxies at $z>10$. These include feedback-free starbursts \citep{Dekel23}, globally enhanced or density-dependent star formation efficiencies \citep{Somerville25}, strongly bursty star formation histories \citep{
Sun23,Gelli24}, evolving or top-heavy IMFs \citep{Yung24,Lu25,Hutter25,Fontanot26,Durrant26}, and changes to dust attenuation \citep{Lu25}. The fact that several physically distinct models can reproduce similar UVLFs reinforces the point that the luminosity function alone is unlikely to discriminate uniquely between them.

The solution that emerges in \shark\ combines several of these ideas, but without introducing any of them specifically to reproduce JWST observations. Stellar feedback remains substantial at high redshift, particularly in dwarf galaxies, and hence the model does not rely on feedback-free star formation. We retain a universal \citet{Chabrier03} IMF. Nor do we impose a globally enhanced star formation efficiency or an explicit redshift dependence in the star formation law. Instead, globally unstable, gas-rich disks trigger starbursts in which molecular gas is converted into stars more efficiently. As these events become increasingly important towards high redshift, they naturally enhance both the luminosities and the stochasticity of the brightest galaxies.

In this respect, the comparison with other SAMs is informative. When disk instabilities are removed, the \shark\ UVLF becomes much closer to those predicted by models such as the Santa Cruz SAM and GAEA, which do not include an equivalent starburst channel associated with globally unstable disks. Conversely, GALFORM also successfully predicted a high abundance of UV-bright galaxies before JWST, although there the combination of bursts, a top-heavy IMF during those bursts, and the treatment of dust plays a central role \citep{Cowley18,Lu25,Elliott26}. These comparisons suggest that rapid, burst-like modes of star formation may be a common ingredient among models that naturally populate the bright end, while the detailed physical origin and stellar populations associated with those bursts differ substantially.

Ultimately, discriminating between these possibilities requires moving beyond the UVLF. In addition to clustering and galaxy sizes, measurements of stellar masses, gas content, metallicities, nebular emission, dust emission and star formation histories can test whether UV-bright galaxies occupy the halo masses and physical states predicted by the different scenarios.

\subsection{A consistent model across cosmic time}

Perhaps the most important aspect of our results is that the physical model employed here was not constructed to reproduce the $z>10$ galaxy population. The parameters of \shark\ were calibrated against the $z\approx0$ SMF, while the prescriptions governing star formation, starbursts and stellar feedback were motivated primarily by observations and theoretical models at substantially lower redshift. Nevertheless, without changing these prescriptions or their parameters, the model provides a reasonable description of the UVLF to $z\sim 17$ (Fig.~\ref{UVLF}), the evolution of the SMF to $z\sim10$ (Fig.~\ref{SMF}), and the CSFRD from $z=0$ to $z\sim 15$ (Fig.~\ref{cosmicsfr}).

This provides an important complement to attempts to determine what changes to galaxy formation physics are required by {\it JWST}. A sufficiently flexible modification can generally improve agreement with a particular high-redshift observable; the more stringent requirement is that it does so without compromising the galaxy population at other epochs. In \shark, the same physical prescriptions naturally produce very different effective modes of galaxy growth as the conditions of galaxies evolve. Starbursts triggered by unstable disks become increasingly important for UV-bright galaxies towards cosmic dawn in \shark, while ordinary disk star formation dominates much of the later galaxy population. The model therefore does not require galaxy formation physics to change abruptly at high redshift: instead, the relative importance of processes already present in the model changes as galaxies move into increasingly extreme physical conditions.

The agreement demonstrated here should not be interpreted as evidence that the detailed prescriptions in \shark\ are uniquely correct. In particular, the treatment of violent disk instabilities and high-redshift dust remain deliberately simplified, and several other models achieve similar UVLFs through different mechanisms. Rather, our results demonstrate that the abundance of UV-bright galaxies observed by the {\it JWST} does not by itself require either new cosmological physics or baryonic prescriptions that operate exclusively in the early Universe. A galaxy formation framework constrained by the later Universe can naturally produce such a population when it contains mechanisms capable of generating sufficiently efficient and stochastic episodes of star formation at early times. The challenge for the next generation of observations is to determine which physical mechanism is responsible for those episodes.

\section{Conclusions}\label{conclusions}

We have presented predictions for the rest-frame UV luminosity function of galaxies at $6\leq z\leq 17$ using the \shark\ semi-analytic model of galaxy formation. The version of \shark\ used here was calibrated against the $z\approx0$ stellar mass function and has not been recalibrated to reproduce JWST observations. Our main conclusions are:

\begin{itemize}
    \item  The fiducial \shark\ model reproduces the observed evolution of the UVLF from $z\sim6$ to the constraints currently available at $z\sim 15-17$, including the abundance of the brightest galaxies at $z>10$ (Fig.~\ref{UVLF}). This agreement is obtained without introducing new redshift-dependent baryonic physics, suppressing stellar feedback, or varying the IMF.

    \item Starbursts triggered by globally unstable disks are critical to this success. At $z\gtrsim10$, most galaxies brighter than $M_{\rm UV}\approx-17$ have UV emission dominated by instability-driven starbursts (Fig.~\ref{UVLFchannels}). Removing disk instabilities suppresses the bright end of the UVLF by up to $\approx 1$~dex at the highest redshifts (Fig.~\ref{UVLFnodiskins}).

    \item The primary effect of these instability-driven starbursts is to increase both the typical UV luminosity and its scatter at fixed stellar and halo mass (Fig.~\ref{UVscaling}). The resulting increase in stochasticity towards the bright, high-halo-mass population is what allows \shark\ to populate the observed bright end of the UVLF. This differs from scenarios in which enhanced stochasticity towards  {\it lower} halo masses produces the UV-bright population.

    \item Enhanced star formation efficiency during bursts is also required (Fig.~\ref{UVLFnodiskins}). Setting the burst efficiency equal to that of ordinary disk star formation produces too few bright UV galaxies even when the instability channel remains active. Importantly, the star formation efficiency enhancement adopted in \shark\ was motivated by observations of starbursts at $z\lesssim 2$ rather than introduced to reproduce {\it JWST} observations.

    \item Dust attenuation has a significant impact on the bright UVLF at $z\lesssim10$, reaching $\sim1.5$~mag at $z=10$ (Fig.~\ref{UVLF}), but becomes progressively less important towards the highest redshifts. The uncertainty associated with extrapolating locally calibrated dust prescriptions into the first few hundred Myr remains an important limitation of the current modelling.

    \item Reliable predictions during cosmic dawn require both sufficient mass and temporal resolution. Low-mass haloes make an important contribution to the faint UVLF and CSFRD, while the rapid evolution of halos and short-lived starbursts require merger trees with sufficiently fine time sampling. For the implementation of \shark\ studied here, our tests indicate that $\gtrsim 70$ snapshots at $z>6$ are required for convergence of the predicted UVLF.
\end{itemize}

Most importantly, the same \shark\ model that reproduces the high-redshift UVLF also provides a reasonable description of the evolution of the stellar mass function from $z=0$ to $z\sim10$ and of the cosmic star formation rate density from $z=0$ to $z\sim15$. The physical prescriptions and parameters are unchanged across this entire interval. Our results therefore show that the abundance of UV-bright galaxies revealed by the {\it JWST} can emerge naturally within a galaxy formation model anchored to the low-redshift Universe. Rather than requiring qualitatively new physics at cosmic dawn, the observations may be revealing a regime in which processes already familiar from later epochs—gas-rich instabilities and highly efficient starbursts—become sufficiently frequent to generate a strongly stochastic UV-bright galaxy population.

The specific implementation of these processes in \shark\ is not unique, and the UVLF alone cannot distinguish between competing physical explanations. Future measurements of galaxy clustering, galaxy sizes and morphologies, gas and dust content, metallicities and star formation histories will therefore all provide parts of the puzzle to establish whether the bright galaxies seen by the {\it JWST} are indeed the short-lived, instability-driven starbursts predicted here.

\section*{Data Availability}

The medi-SURFS and micro-SURFS halo and subhalo catalogues and corresponding merger trees used in this work can be accessed from:
\newline \url{https://data.pawsey.org.au/download/WAVES/microSURFS_L40_N512_HBTcatalogues.tar.gz} 

\noindent and 

\noindent
\url{https://data.pawsey.org.au/download/WAVES/mediSURFS_L210_N1536_HBTcatalogues.tar.gz}

\noindent The data associated with Hi-SURFS and L800 can be shared upon reasonable request. \shark\ is a public code and the source and python scripts used to produce the plots in this paper can be found at \url{https://github.com/ICRAR/shark/}.

\section*{Acknowledgements}

We thank Jorge Zavala and Hakim Atkins for sharing their data compilations and the constructive discussions. This work was supported by resources provided by The Pawsey Supercomputing Centre with funding from the Australian Government and the Government of Western Australia. We acknowledge the Virgo Consortium for making
their simulation data available (for L800). ChatGPT (OpenAI) was used to assist with language editing, manuscript structure, and literature searches. The authors take full responsibility for the scientific content of the manuscript.

%%%%%%%%%%%%%%%%%%%%%%%%%%%%%%%%%%%%%%%%%%%%%%%%%%



%\bibliographystyle{mnras}

\bibliographystyle{mn2e_trunc8}
\bibliography{SharkSFz10}

%%%%%%%%%%%%%%%%%%%%%%%%%%%%%%%%%%%%%%%%%%%%%%%%%%

%%%%%%%%%%%%%%%%% APPENDICES %%%%%%%%%%%%%%%%%%%%%

\appendix

\section{\shark\ calibration}\label{newcalibration}

A key difference between this work and \citet{Lagos24} is that here we adopted the methodology introduced by \citet{Chandro-Gomez25a} to minimise the impact of numerical artefacts that typically impact merger trees and halo catalogues. This necessarily requires a recalibration of the model parameters as this new methodology changes the scaling of some of the processes from using subhalo to host halo properties, for instance on the gas cooling treatment (see section~4.1 in \citealt{Chandro-Gomez25a}). Specifically, these changes include:

\begin{enumerate}
    \item Non-physical central-satellite hierarchy swaps: to avoid this effect impacting the evolution of galaxies, galaxy properties for central subhaloes are defined using the properties of their host haloes. For satellites and orphan galaxies, their global properties are defined using the properties of the subhalo at infall (just before it becomes a satellite), except for environmental processes such as ram-pressure and tidal stripping, where the current subhalo properties are used. To run \shark\ with this implementation, the user needs to add the following option to the parameter file:\newline \noindent {\tt apply$_{-}$fix$_{-}$to$_{-}$mass$_{-}$swapping$_{-}$events = true}

    \item Sudden appearance of massive subhaloes at late cosmic times: a subhalo-progenitor connection is flagged as non-physical if a subhalo is born with an unrealistically large number of particles. For flagged subhaloes, we break the current progenitor-subhalo link and connect the progenitor to another subhalo based on the smoothness of the subhalo mass gained or lost. In this case, the following option must be added to the parameter file to activate the modification: \newline {\tt apply$_{-}$fix$_{-}$to$_{-}$massive$_{-}$transient$_{-}$events = true}
\end{enumerate}

Apart from the parameters related to avoiding the impact of numerical issues, Table~\ref{newparams} shows the values of the parameters that are different from those adopted in \citet{Lagos24}, also recalling the value adopted there for that parameter and the physical mechanism that it affects.

\begin{table}
    \centering
        \caption{List of parameters that were modified in the process of calibration of \shark\ in this work relative to those adopted in \citet{Lagos24}, which are provided in parenthesis for reference. We also list the physical process in which each of these parameters appear.}
    \label{newparams}
    \begin{tabular}{c|c|c}
    \hline
    \hline
    Parameter & Value adopted & Equation \\
    \hline
    \hline
     Process & AGN Feedback& \\
    \hline
         {\color{black}$\kappa_{\rm jet}$} &  {\color{black}$0.0992$ ($0.023$)} & Eq.~32 in\\
         & & \citet{Lagos24}\\
 \hline
 \hline
 Process & Ram-pressure Stripping & \\
 \hline
 $\alpha_{\rm cold}$ & $100$ (1) & Eq.~2 in\\
 & & \citet{Oxland26}\\
 \hline
 \hline
Process & Stellar Feedback & \\
\hline
{\color{black}$z_{\rm P}$} & {\color{black}$0.134$ (0.2)} & {\color{black}Eq.~\ref{vel_power}}\\
\hline
\hline
Process & Reincorporation & \\
\hline
{\color{black}$M_{\rm norm}$} & {\color{black}$3.411 \times 10^{10}\,\rm M_{\odot}$ ($1.383\times 10^{11}\,\rm M_{\odot}$)} & Eq.~30 in \\
& & \citet{Lagos18c}\\
\hline
\hline
Process & Disk Instabilities& \\
\hline
$\epsilon_{\rm disc}$ & $0.986$ (1) & Eq.~35 in\\
& & \citet{Lagos18c}\\
$f_{\rm int}$ & $4.735$ (2) & Eq.~48 in \\
& & \citet{Lagos18c}\\
\hline
    \end{tabular}
\end{table}

%In addition to the parameters changes, to run \shark\ with the implementation of the fixes to the numerical artefacts affecting merger trees and halo/subhalo catalogues that were introduced by \citet{Chandro-Gomez25a}, the user needs to add to the parameter file the options:  
%\newline {\tt apply$_{-}$fix$_{-}$to$_{-}$massive$_{-}$transient$_{-}$events = true}
%\newline \noindent {\tt apply$_{-}$fix$_{-}$to$_{-}$mass$_{-}$swapping$_{-}$events = true}.

\section{The impact of the adopted dust-to-metal scaling relation on the UVLF at $z>6$}\label{dustatt}

The calculation of dust attenuation in \shark\ depends explicitly on the dust surface density of galaxies (see \S~\ref{dustsec}). In \shark, we do not explicitly track the formation, destruction and growth of dust in the ISM, and instead, we adopt a dust-to-metal scaling relation informed from the observations of \citet{Remy-Ruyer14}. Because of the uncertainty associated with this scaling, in this Appendix we test the impact the adopted scaling has on the resulting UVLF. We remind the reader that this only impacts the attenuated UVLF. 

\begin{figure}
\begin{center}
\includegraphics[trim=2mm 2mm 2mm 2mm, clip,width=0.49\textwidth]{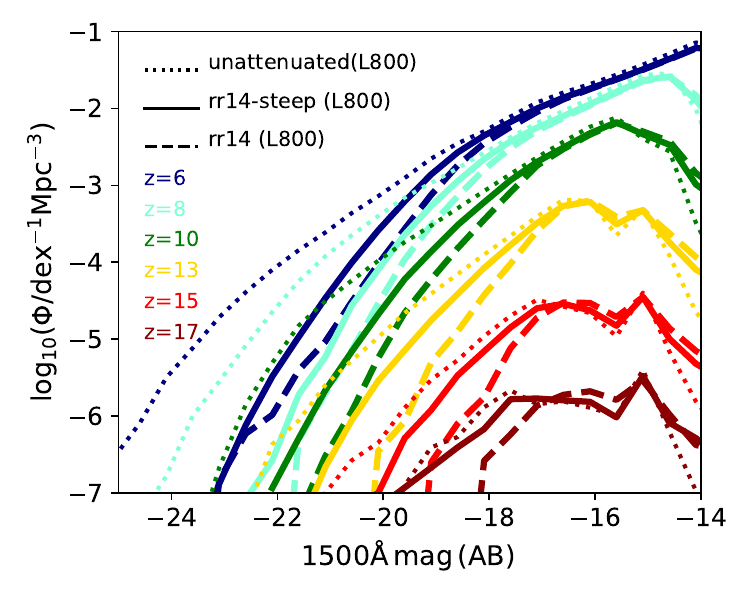}
\caption{The rest-frame unattenuated UV luminosity function (UVLF) at $z=6,\, 8,\,10,\,13,\,15,\,17$, as labelled, in \shark\ when run over the L800 box. We show three versions of the UVLF: the unattenuated LF (dotted lines), the resulting UVLF after attenuation is considered and adopting the dust-to-metal scaling relations ``rr14'' (dashed lines) and ``rr14-steep'' (solid lines) our fiducial run (solid lines). In the fiducial runs shown in Figs.~\ref{UVLF}~and~\ref{UVLFnodiskins} we use the ``rr14-steep'' scaling.} 
\label{UVLFnodiskins2}
\end{center}
\end{figure}

Fig.~\ref{UVLFnodiskins2} shows the UVLF at $z\ge 6$ in \shark\ using the L800 box (see Table~\ref{tab:sims}) under the two dust-to-metal scaling relations introduced in \S~\ref{dustsec}. Overall, we see that the ``rr14-steep'' scaling, which is the default one adopted for the UVLF at $z\ge 6$, the level of attenuation in the UV is quite minimal at $z=17$, while becoming increasingly more important with decreasing redshift. Conversely, the ``rr14'' scaling still produces a significant attenuation at the bright end of the UVLF even at $z=17$, with the brightest galaxies seeing attenuations of $\approx 1-1.5$~mag at these high redshifts. By $z=6$, however, both dust-to-metal scalings yield similar attenuated UVLF. 

The difference at $z=17$ is due to most of these bright galaxies having metallicities between $\rm log_{10}(Z/Z_{\odot})=[-0.59,-0.16]$, which mark the transition between a constant dust-to-metal ratio and a decrease with decreasing metallicity in the ``rr14'' and ``rr14-steep'' scalings (see Eqs.~\ref{eq.mdust}~and~\ref{eq.mdust2}). In reality, any delay in dust formation relative to the metal enrichment of the ISM would lead to a smaller dust-to-metal ratio in the early Universe, which is why the steeper scaling of Eq.~\ref{eq.mdust2} performs better against the observed UVLF in the very early Universe.
%\input{AppendixSpin}
%%%%%%%%%%%%%%%%%%%%%%%%%%%%%%%%%%%%%%%%%%%%%%%%%%

% Don't change these lines
\bsp	% typesetting comment
\label{lastpage}
\end{document}